\documentclass[manuscript]{aastex}

\slugcomment{To be submitted to the Astronomical Journal}

\shorttitle{Extra-tidal tails of globular clusters}
\shortauthors{Chun et al.}

\begin{document}

\title{A wide-field photometric survey for extratidal tails around five metal-poor globular clusters
in the Galactic halo
\footnote{Based on observations carried out at the Canada-France-Hawaii
        Telescope, operated by the National Research Council of Canada,
        the Centre National de la Recherche Scientifique de France,
        and the University of Hawaii.
        This is part of the Searching for the Galactic Halo project using the CFHT,
        organized by the Korea Astronomy and Space Science Institute.}}
\author{Sang-Hyun Chun\altaffilmark{2}, Jae-Woo Kim\altaffilmark{3}, Sangmo T. Sohn\altaffilmark{4,7}, 
Jang-Hyun Park\altaffilmark{5}, Wonyong Han\altaffilmark{5}, Ho-Il Kim\altaffilmark{5}, 
Young-Wook Lee\altaffilmark{2,4}, Myung Gyoon Lee\altaffilmark{6}, Sang-Gak Lee\altaffilmark{6}, 
and Young-Jong Sohn\altaffilmark{2}}
\altaffiltext{2}{Department of Astronomy, Yonsei University, Seoul 120-749, Korea, sohnyj@yonsei.ac.kr}
\altaffiltext{3}{Institute for Computational Cosmology, Department of Physics, Durham University, 
South Road, Durham DH1 3LE, UK}
\altaffiltext{4}{Center for Space Astrophysics, Yonsei University, Seoul 120-749, Korea}
\altaffiltext{5}{Korea Astronomy and Space Science Institute, Daejeon 305-348, Korea}
\altaffiltext{6}{Astronomy Program Department of Physics and Astronomy, Seoul National University, 
Seoul 151-742, Korea}
\altaffiltext{7}{Current address: Space Telescope Science Institute, 3700 San Martin Drive, 
Baltimore, MD 21218}

\begin{abstract}

Wide-field deep $g^{'}r^{'}i^{'}$ images obtained with the Megacam of the 
Canada-France-Hawaii Telescope (CFHT) are used to investigate the spatial configuration of stars 
around five metal-poor globular clusters M15, M30, M53, NGC 5053, and NGC 5466, 
in a field-of-view $\sim3^{\circ}$. 
Applying a mask filtering algorithm to the color-magnitude diagrams of the observed stars,
we sorted cluster's member star candidates that are used to examine the characteristics 
of the spatial stellar distribution surrounding the target clusters.
The smoothed surface density maps and the overlaid isodensity contours indicate that
all of the five metal-poor globular clusters exhibit
strong evidence of extratidal overdensity features over their tidal radii,
in the form of extended tidal tails around the clusters.
The orientations of the observed extratidal features 
show signatures of tidal tails tracing the clusters' orbits, inferred from their proper motions,
and effects of dynamical interactions with the Galaxy.
Our findings include detections of a tidal bridge-like feature
and an envelope structure around the pair of globular clusters M53 and NGC 5053.
The observed radial surface density profiles of target clusters have 
a deviation from theoretical King models, 
for which the profiles show a break at $0.5\sim0.7r_t$, 
extending the overdensity features out to $1.5\sim2r_t$.
Both radial surface density profiles for different angular sections and 
azimuthal number density profiles confirm the overdensity features of 
tidal tails around the five metal-poor globular clusters.
Our results add further observational evidence that the observed metal-poor
halo globular clusters originate from an accreted satellite system,
indicative of the merging scenario of the formation of the Galactic halo.

\end{abstract}


\section{INTRODUCTION}

In the current cold dark matter cosmology,
the favorably accepted galaxy formation paradigm is that galaxies like the Milky Way 
were assembled from a series of accretion events 
involving lower mass subunits, in a fashion similar to galaxy cluster formation
~\citep{Whi78,Blu84,Dav85,Tor97,Tor98a,Ghi98,Tor98b,Ghi00}.
If these favored hierarchical models are correct, 
then the stellar halo of the Galaxy should form 
through tidal disruption and accretion of numerous low-mass fragments 
such as dwarf satellite galaxies~\citep[e.g.,][]{Joh98,Kly99,Moo99,Moo06,Bul01,Bul05,Aba06}. 
These merging events may leave traces of large-scale stellar substructures 
in the outer Galactic halo, and remnant stellar systems
of the parent fragments with a long dynamical disruption time scale.

Since the classical ~\citet{Sea78} scenario of the formation of
the Galaxy was first suggested, there has been more observational evidence
to support the accretion scenario that the halo population of the Milky Way 
and globular clusters originated through the gradual merging and 
disruption of fragments.
This evidence includes (1) a large-scale stellar substructure through the Galactic halo,
(2) a feature of multiple populations in a globular cluster 
that is indicative of the remnant nuclei of a merged dwarf spheroidal galaxy, 
(3) an extended tidal feature in a spatial stellar density distribution,
such as tidal tails and halos around a globular cluster, which might trace the cluster's orbit,
and (4) a spatially aligned configuration of a group of several globular clusters.
Consequently, globular clusters in the Galactic halo and their surroundings, 
as the local manifestations of the hierarchical Galaxy formation,
are the best places to look for the substructure of stellar distribution 
in the Galactic outer halo.

For the first note of large-scale stellar substructures through the Galactic halo,
~\citet{Lyn82} indicated that several dwarf
spheroidal galaxies and outer halo clusters lie near two great circles passing
close to the Galactic poles, i.e., the Magellanic stream and the Fornax-Leo-Sculptor
stream. The spatially associated distribution of dwarf galaxies is interpreted
as a result of the accretion of parent satellite galaxies
which are considered the external fragments.
Furthermore, recent large-scale surveys have uncovered several complex substructures
of the stellar Galactic halo~\citep[e.g.,][]{Yan00, Yan03, Ive00, New02, New09, Mar04, 
Roc04, Mar05, Gri06, Bel06a, Jur08}.
The most striking example in this category is the discovery of 
the trails of stellar debris associated with
the Sagittarius dwarf galaxy ~\citep{Iba94,Iba95,Iba97,Iba01a,Maj03,New03}. 

Stellar population studies of globular clusters
have suggested that the brightest globular clusters in the Galaxy
are actually the remnant nuclei of dwarf spheroidal galaxies. 
Indeed, ~\citet{Lee99} detected multiple stellar populations of different age and 
metallicity in $\omega$ Centauri, the most massive globular cluster in the Galaxy. 
This is reminiscent of the Sagittarius dwarf galaxy, 
which includes M54 as its nucleus, the second most massive globular cluster in the Galaxy.
Extending previous suggestions, ~\citet{van00} suggested that the majority of metal-poor
young globular clusters in the outer halo originated from dwarf spheroidal galaxies,
and it would be worthwhile to find their parent galaxies.
Recently, ~\citet{Lee07} found that
globular clusters with an extended horizontal-branch (EHB) 
would be kinematically decoupled from the normal clusters, 
and, that these are the relics of the first building blocks of
the Galaxy formation.

If present Galactic globular clusters formed within larger stellar systems,
they are likely surrounded by extratidal halos and tails 
made up of stars that were tidally stripped from their parent systems.
Also, they would have lost the majority fraction of the initial mass 
due to their internal and external dynamical effects,
such as tidal heating and stripping~\citep{Ost72, Agu88, Kun95,Gne97,Mur97,Gne99,Bau03}.
This information suggests that surroundings around globular clusters provide 
an excellent example of such a structure,
and as a result we might expect to figure out them by wide-field
deep photometry~\citep[e.g.,][]{Mart04}.
In the first attempt to find such a tidal feature around globular clusters,
~\citet{Gri95} examined the stellar density distribution around
12 globular clusters and found that 
many globular clusters have surface density profiles significantly exceeding
the prediction of formal King models and also 
extending outside the tidal radius by extratidal stars.
Until now, signatures of the extratidal overdensity feature have been found in 
over 30 globular clusters both in the Galaxy and in M31~\citep{Gri96,Leh97,Leo00,Tes00,
Sie01,Lee03,Soh03,Lau06}.
Recent star count surveys have revealed tidal structures stretching
many degrees beyond the cluster tidal radius,
for which the most spectacular examples include
Palomar 5 ~\citep{Ode01,Ode03,Ode09,Roc02,Gri06} and NGC 5466 ~\citep{Bel06b,Gril06}
with tails extending as much as $22^{\circ}$ and $45^{\circ}$, respectively.
As a consequence of these previous studies, it has been apparent that tidal tails
of globular clusters are likely aligned with the cluster's orbit because 
the debris produced by a dissolving globular cluster will have similar 
orbital parameters as the cluster itself.
Recently, ~\citet{Ode09} presented kinematics of debris around Palomar 5 to suggest
that the cluster's orbit in the sky is not exactly aligned with its tidal tails.
Here, we also note that ~\citet{Koc04} found that the luminosity function (LF) 
in the tails of Palomar 5 agrees well with the LF in the outer region of the cluster
but differs from the LF in the cluster's core.
On the theoretical side, many numerical simulations have reproduced 
a structure of tidal tails around globular clusters~\citep{Deh04,diMa05,Lee06,Fel07,Kup08}. 
Indeed, the cluster-galaxy interaction can cause stars to escape
the cluster's gravitational potential, 
and then the unbounded stars remain in the vicinity of the cluster 
for several orbital periods to become tidal debris-forming tails
in front of and behind the cluster's orbit~\citep{Com99,Yim02,Cap05,Mon07}.

The spatial alignment of the globular cluster system might also play an important role in 
untangling the possibility that the Galactic halo may have formed from the accretion 
of dwarf galaxies. For a statistically significant instance, 
the Sagittarius dwarf spheroidal has its own globular cluster system associated 
with several halo clusters beyond a galactocentric distance $R_{GC} \ge 10$ kpc
~\citep{Bel03,Tau04,Sbo05}.
This attests to the fact that even minor mergers play a key role 
in the formation and structure of the Galaxy.
Furthermore, ~\citet{Yoo02} found that a group of the seven lowest metallicity
([Fe/H] $<$ -2.0) clusters (i.e., NGC 5053, NGC 5466, M15, M30, M53, M68, and M92,
classified as Oosterhoff group II-b)
displays a planer alignment in the outer halo perpendicular 
to the line joining the present position of the Sun and the Galactic center.
They also suggested that these metal-poor clusters were
originally born in the Large Magellanic Cloud (LMC) and have recently been captured by the Galaxy
through the Magellanic plane.
Indeed, the configuration of the Milky Way with the stream of clusters
and satellites in the Magellanic plane closely resembles
the stellar stream in the halo of M31, which lies along the satellites
M32 and NGC 205~\citep{Iba01b}. 
Of the seven low-metallicity clusters in the Magellanic plane ~\citep{Yoo02}, 
NGC 5466 has been the subject of extensive investigations from wide-field photometric analyses 
of the SDSS data ~\citep{Bel06b,Gril06} 
to show a feature of extended extratidal tails outside the tidal radius along the cluster's orbit. 
The other clusters in the Magellanic plane have not yet shown any strong observational evidences of 
extratidal tails along the clusters' orbits.
Instead, hints of a weak evidence of extratidal halo structure outside the tidal radius 
have been reported for three clusters, 
i.e., M15 ~\citep{Gri95}, M68 ~\citep{Gri95}, and M92 ~\citep{Tes00, Lee03}.

In this paper, we investigate the two-dimensional distribution of stars
around five clusters (M15, M30, M53, NGC 5053, and NGC 5466) of 
the seven extreme metal-poor globular clusters of the
Oosterhoof group II-b ~\citep{Yoo02} and we report the spatial structure of 
the stellar distribution in the vicinity of the clusters.
Wide-field $g^{'}r^{'}i^{'}$ imaging photometric data for the five globular clusters 
have been secured from 
the observations of the Megacam attached to the Canada-France-Hawaii Telescope (CFHT).
Table~\ref{posi} contains basic data for the five target globular clusters.
Note that the other two low-metallicity clusters M68 and M92 have not been observed
at the present observing runs.
Section 2 presents the observations, data reduction procedure, 
and photometric measurements of the resolved stars.
The statistical mask filtering method for the photometric selection of clusters' member stars
is presented in Section 3.
In Section 4, we investigate extratidal features around the target clusters
by using two-dimensional stellar density maps, radial surface density profiles,
and azimuthal number density profiles.
The results are discussed and summarized in Section 5.

\section{OBSERVATIONS AND DATA REDUCTION}

\begin{figure}
\centering
\includegraphics[width=0.8\textwidth]{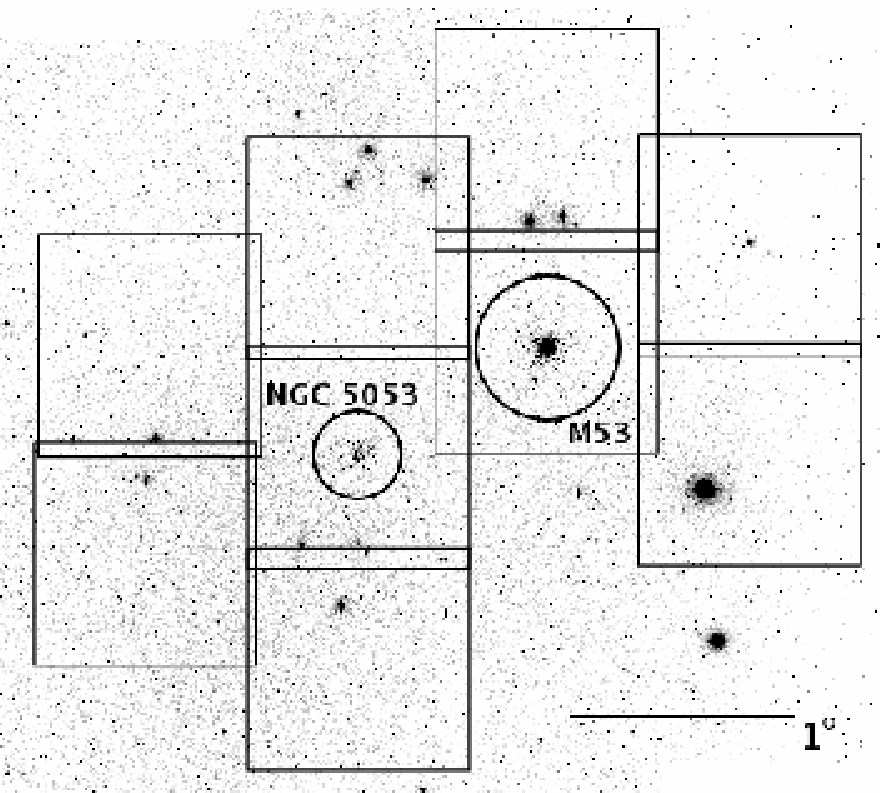}
\caption{The Digitized Sky Survey image for the field of M53 and NGC 5053. 
Boxes represent the observed nine Megacam fields.
Circles indicate the tidal radii for each cluster.
North is up and east to the left. 
\label{obs}
}
\end{figure}

The wide-field photometric imaging data for M15, M30, M53, NGC 5053, and NGC 5466
were taken in Queued Service Observing (QSO) of the 3.6 m CFHT.
The detector in the Megacam is a mosaic of thirty-six $2048\times4612$
pixels$^2$ CCDs, that together cover an area of $\sim 1^{\circ}\times1^{\circ}$ 
wide-field-of-view with a plate scale of $0.187^{''}$ pixel$^{-1}$.
We covered nine and eight Megacam observing fields for the surroundings of 
two clusters, M15 and NGC 5466, and of one cluster, M30, respectively.
For each cluster, one field was centered to imbed the cluster itself,
while the other fields outside the object were chosen so as to cover
a total field of view of $\sim 3^{\circ}\times3^{\circ}$.
Since the tidal radii of the clusters are in the range of $14^{'}$ $\sim 22^{'}$,
a set of these observations provided sufficiently large spatial coverage
for our purpose of studying the stellar density distribution and tidal features
in the vicinity of the globular clusters.
The other two clusters, M53 and NGC 5053, compose a close neighbor pairing system
with an angular separation of $\sim 1^{\circ}$ on the sky, and the two clusters
have nearly the same distance modulus.
Note that the close proximity of the two clusters is very unusual
in the outer halo of the Galaxy.
In order to search for direct observational evidence of a tidal interaction
between the two clusters and tidal features around each cluster,
we observed nine Megacam fields in a pattern as shown in Figure~\ref{obs},
covering both of the clusters.

The imaging data for the target clusters were taken in three
SDSS filters, $g^{'}$, $r^{'}$, and $i^{'}$, respectively.
Single exposures were taken for each filter without a dither pattern.
The exposure times to be used were determined so as to achieve a limiting magnitude of
$\sim 3$ magnitudes below the main sequence turnoff of each cluster, i.e., $g^{'} \sim 23$ 
mag. Table~\ref{log} presents the journal of the observations.
Stars have $0.6-1.0$ arcsec FWHM in the final processed images, depending on the filter.

The removal of instrumental signatures from the raw images was done with the ELIXIR
processing pipeline at the CFHT~\citep{Mag04}. This processing
includes bias subtraction, flat fielding, and the subtraction of a fringe frame.
Also, the pre-analysis of the ELIXIR produced 36 CCD images of a given mosaic
with the same zeropoint and magnitude scale. A typical value of the residuals
across a mosaic is $\sim 0.0086$ mag.

The photometric measurements were made with the point-spread-function (PSF)
fitting programs DAOPHOT II and ALLSTAR~\citep{Ste87,Ste88}.
Analysis was done on a single-chip basis for all exposures of the 36 CCD chips on the mosaic.
A PSF varying quadratically through positions was constructed
using $\sim 100-200$ bright and isolated stars for a heavily crowded image and
$\sim 25-50$ stars for a less crowded image.
The quality of each PSF has been improved by removing neighboring stars
around selected PSF stars and reconstructing the PSF iteratively.
Instrumental magnitudes for each star based on the determined PSF
were then measured by the ALLSTAR process.
Aperture correction was applied by using
the growth-curve analysis package DAOGROW~\citep{Ste88,Ste90}
to compensate for the missing light lying outside of the PSF tail
for every star in each frame.
After the aperture correction,
DAOMATCH and DAOMASTER~\citep{Ste92} were used to match stars
on each CCD chip, and coordinate offsets for each field were estimated
from the positional data in each frame.
We then converted all the positional data of each chip on a mosaic
into a single coordinate system.
Finally, the positional data for stars in the observed Megacam fields were transformed 
into the equatorial coordinate system, 
by using the USNO-B catalog ~\citep{Mon03} for bright stars in the fields.

The measured instrumental magnitudes were photometrically calibrated using
the zeropoints that were computed from standard star observations
during each Megacam run.
The calibrated magnitudes were finally converted into a standard SDSS
photometric system, as defined by ~\citet{Smi02}.
Individual stars with poorly determined photometric measurements,
as reported by ALLSTAR, were removed
to limit the number of spurious detections for each cluster.
Objects with $\chi$ and SHARP values that markedly deviated
from the values for the majority of objects were rejected, and
this removed objects near the faint limit of the data and
extended objects such as faint galaxies.
In order to correct the interstellar extinction,
we derived the individual extinction value for each detected star
from the maps of~\citet{Sch98}, and these values have been subtracted from the
observed magnitude. 
Finally a total of 239 600, 100 069, and 69 901 stars were detected
for each $\sim3^{o}\times3^{o}$ field of clusters M15, M30, and NGC 5466, respectively.
For nine Megacam fields of M53 and NGC 5053, we detected
79 075 stars in all three $g^{'}r^{'}i^{'}$ images.

\section{PHOTOMETRIC SELECTION OF THE CLUSTER STARS}

A feature of spatial stellar distribution around globular clusters represents 
the sum of the unbound stars released by the cluster and the field stars.
Therefore, to trace the overdensity features around the clusters in detail,
one needs to minimize the number of contaminating field stars in a given sample
and to enhance the contrast between the cluster and the field.
In the first statistical attempt to discriminate field stars, 
~\citet{Gri95} introduced a polygonal mask method in a color-magnitude (C-M) space
in which only field stars with colors and magnitudes that resemble those of cluster stars
are counted to optimize the ratio of the cluster stars to field stars.
However, the defined mask in a C-M space does not provide the optimal method to map 
the spatial stellar distribution around sparsely populated globular clusters 
because many bins with a small number of stars in a C-M space can be 
statistically excluded from the mask. 
To reduce the disadvantage of poor statistics,
~\citet{Ode01} introduced a new method to compress the photometric data in a C-M space 
by constructing new color indices from which the weighted sum of field stars plus 
cluster stars yields the best approximation of the observed total C-M distribution.
In the present study, we have applied the same method of ~\citet{Ode01} to define
a C-M mask with  new color indices $c_1$ and $c_2$  and $g^{'}r^{'}i^{'}$ magnitudes
for detected stars in the vicinity of each cluster.

First, we defined two orthogonal color indicies $c_1$ and $c_2$ from a linear distribution
of stars in a ($g^{'}-r^{'}$, $r^{'}-i^{'}$) color-color diagram.
Indeed, the choice of the indicies is such that the main axis of
the almost one-dimensional locus of cluster stars in the color-color plane
lies along the $c_1$ axis, while the $c_2$ axis is perpendicular to it
~\citep[e.g.,][]{Ode01,Ode03}.
Equation~(\ref{eq:rela1}) presents general forms of the rotational transformation of the
($g^{'}-r^{'}$, $r^{'}-i^{'}$) plane to the coordinate system of ($c_1$, $c_2$).
Here, the systematic variation of color with magnitude
is mostly contained in $c_1$, while variations in $c_2$ are mostly due to
the observational photometric errors, as noted in ~\citet{Ode01}.
Transformation coefficients $a$ and $b$ for each cluster were determined
by using the least-squared fit to the data, which are listed in Table 3.
Assuming that two clusters M53 and NGC 5053 have very similar color-magnitude diagrams(CMDs)
with almost identical metallicity and age ~\citep[e.g.,][]{Hea91,Rey98}, and
nearly the same distance modulus,
we estimated a set of $c_1, c_2$ and $a, b$ for all stars detected in both of the clusters.

\begin{eqnarray}
\label{eq:rela1}
c_1=a(g^{'}-r^{'})+b(r^{'}-i^{'}), \\
c_2=-b(g^{'}-r^{'})+a(r^{'}-i{'}) \nonumber
\end{eqnarray}

In ($c_2, i^{'}$) space of stars located in the central region of the clusters,
we preselected the sample in $c_2$ by discarding all stars with 
$|c_2|>2\sigma_{c_2}(i^{'})$, where $\sigma_{c_2}(i^{'})$  is the photometric 
errors in $c_2$ for stars with magnitude $i^{'}$.
The radii of the cluster's central regions were empirically defined
in the unit of half-mass radius ($r_h$) or core radius ($r_c$) of the cluster
in order to find appropriate color-magnitude sequences of stars in each cluster.
The heavy lines in the leftmost panels in Figure 2 give $2\sigma_{c_2}(i^{'})$ limits
for stars in each cluster.
For a system of M53 and NGC 5053, we selected a central region of the clusters
as the sum of $3r_h < r \leq 8r_h$ for M53 and $r_c < r \leq 4r_c$ for NGC 5053.

\begin{figure*}
\centering
\includegraphics[width=0.6\textwidth]{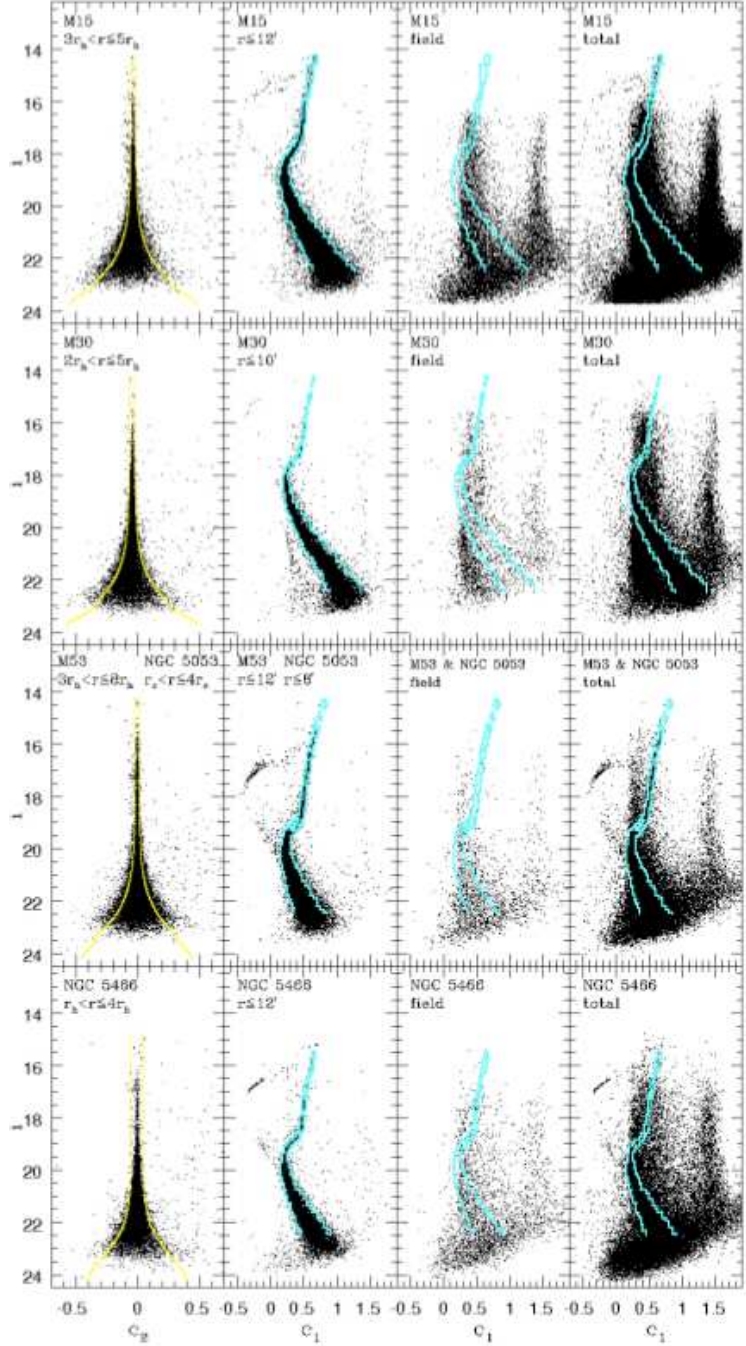}
\caption{The $(c_2,i^{'})$ and $(c_1,i^{'})$ color-magnitude diagrams of stars
in the fields of five target globular clusters. 
From left to right, the first panel is the $(c_2, i^{'})$ plane for stars 
in the empirically determined central regions of each cluster,
the second to fourth panels are the $(c_1, i^{'})$ planes for stars 
in the clusters central area, in the assigned backgournd fields, and in the total 
observed field, respectively. The lines in the $(c_2, i^{'})$ plane indicate $2\sigma_{c_2}$ 
limits at $i^{'}$ magnitudes due to photometric error.
The lines in the $(c_1,i^{'})$ planes are the selected subgrid area determined through the
mask filtering method.\label{cmd1}}
\end{figure*}

\begin{figure*}[t]
\centering
\includegraphics[width=0.8\textwidth]{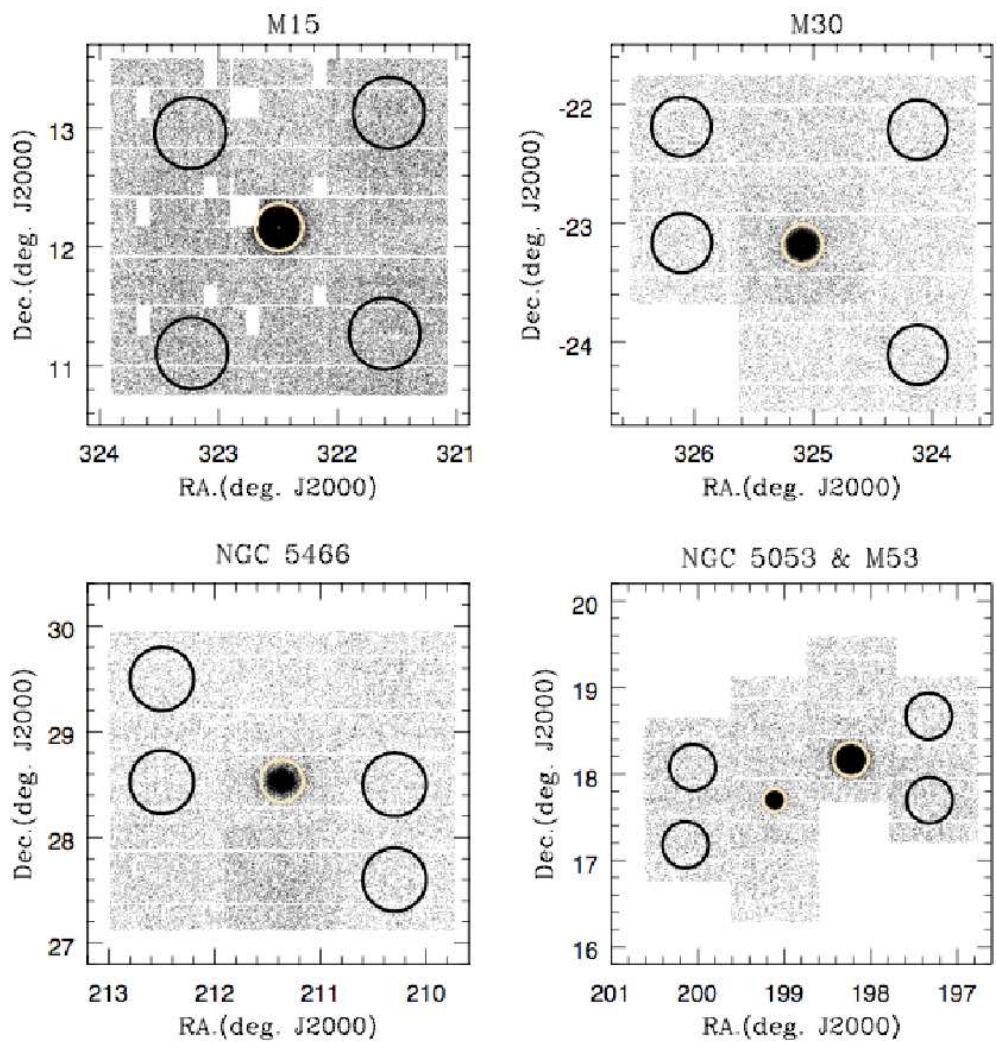}
\caption{
Spatial distribution of all stars measured in the vicinity of five globular clusters. 
Circles centered on each cluster indicate the cluster's central areas
in which the number ratio of stars enclosed in $2\sigma_{c_1}(i^{'})$ and those
outside of $2\sigma_{c_1}(i^{'})$ in the $(c_1,i^{'})$ plane in Figure~\ref{cmd1}
has a maximum value. The other four circles in each field indicate the background areas
for guaging the field star contamination.
~\label{select}}
\end{figure*}

For stars preselected in the $(c_2,i^{'})$ plane, we then applied the
C-M mask filtering method ~\citep{Gri95} in the $(c_1,i^{'})$ plane,
from which we eventually extracted a sample of stars to trace their spatial configuration
around the clusters.
We first examined the distribution of the cluster population in the $(c_1,i^{'})$ space
by means of preselected stars in the cluster central area and
the background field areas.
The cluster central area was assigned so as to have
the largest number ratio of stars enclosed $2\sigma_{c_1}(i^{'})$ and those
outside $2\sigma_{c_1}(i^{'})$ in the $(c_1,i^{'})$ plane
for all preselected stars within the radius of the circle.
Here, $\sigma_{c_1}(i^{'})$ is an estimated standard deviation of
$i^{'}$ magnitudes of stars at $c_1$.
The second from the left panels in Figure~\ref{cmd1} show
$(c_1,i^{'})$ CMDs of stars within the radii of
the determined central areas of the clusters, which are
denoted by circles centered on each cluster in Figure~\ref{select}. 
We also selected four background field areas in circles
at sufficiently distant parts from the cluster.
For three isolated target clusters, M15, M30, and NGC 5466, 
the sum of the background field areas
is nine times those of the determined cluster's area, as shown in Figure~\ref{select}. 
Note that, as shown in the bottom-right panel of Figure~\ref{select},
the background areas for a field of M53 and NGC 5053
were empirically selected to have a total area five times larger than
the sum of central areas of the two clusters.
The third and forth panels of Figure~\ref{cmd1}, from left to right, show
the $(c_1,i^{'})$ CMDs of the stars in the selective background regions
and in the total survey region  for each cluster, respectively.

Next, we compared the ($c_1$,$i'$) CMD of stars in the selective cluster
central region with that of stars in the assigned background regions,
from which we select areas in the ($c_1$,$i'$) plane with high
significance about the expected true number of cluster stars.
This procedure was done by subdividing the C-M plane into small subgrid elements
with a 0.025-mag width in $c_1$ and a 0.12-mag height in $i'$.
Then, we measured the number of stars
in each subgrid element of cluster area $n_{cl}(c_1, i^{'})$
and those of background fields $n_f(c_1, i^{'})$, for which we
assumed that the distribution of background field stars on the C-M plane does
not vary across the observed field.
The local signal-to-noise ratio $s(c_1,i^{'})$ in each subgrid element was 
then calculated by Equation~(\ref{eq:sn1}), where  $g$ is simply the ratio of
the area of the cluster region to that of the background field.

\begin{eqnarray}
\label{eq:sn1}
s(c_1,i^{'})=\frac{n_{cl}(c_1,i^{'})-gn_f(c_1,i^{'})}{\sqrt{n_{cl}(c_1,i^{'})+g^2n_f(c_1,i^{'})}}
\end{eqnarray}

From the array $s$, we obtained an optimal mask envelope
in the $(c_1,i')$ plane by setting a threshold $s_{lim}<s_{max}$ (maximum of $s$)
and identifying the subgrid elements of $s \geq s_{lim}$.
In order to do this, the elements of $s(c_1,i^{'})$ were sorted into a series of
gradually decreasing thresholds over the one-dimensional index $k$.
From the determined subgrid element with the highest signal-to-noise ratio ($s_{max}$),
a cumulative number of stars in the cluster area $N_{cl}(a_k)=\sum_{l=1}^kn_{cl}(l)$
was then counted through a progressively larger area $a_k=ka_l$,
where $a_l$ is the area of a single element of the C-M plane.
The cumulative star counts for the background field
$N_f(a_k)=\sum_{l=1}^kn_f(l)$ were calculated in the same manner as for the
cluster area. The cumulative signal-to-noise ratio $S(a_k)$ was then determined by 
Equation~(\ref{eq:csn1}):

\begin{eqnarray}
\label{eq:csn1}
S(a_k)=\frac{N_{cl}(a_k)-gN_f(a_k)}{\sqrt{N_{cl}(a_k)+g^2N_f(a_k)}}
\end{eqnarray}

$S(a_k)$ reaches a maximum value
for a particular subarea of the C-M plane, and the corresponding
value of $s(c_1,i^{'})$ is set to a threshold, i.e., $s_{lim}$.
Now, the filtering mask area is chosen by selecting subgrid elements
with larger $s(c_1,i^{'})$ values than the determined $s_{lim}$.
The heavy lines in ($c_1,i^{'}$) CMDs in Figure \ref{cmd1} 
represent the boundaries of the selected subgrid elements.
The faint limits of the filtering mask areas were assigned
to be $i^{'}=22.0$ for NGC 5466 and $i^{'}=22.5$ for the other clusters
to avoid biases caused 
by the poor completeness ($\sim70\%$) of the photometry and 
by the variable seeing conditions over the surveyed sky areas during the observations.
Finally, the entire sample of stars in the determined filtering mask area
was considered in the subsequent analyses to examine 
the spatial configuration around the target clusters.

\section{SPATIAL CONFIGURATION OF STARS}

In this section, we present and discuss the analytic results related to 
the spatial configuration of stars in the vicinity of target clusters individually 
for each cluster.
Details of the extended tidal features with significant overdensities of stars 
over the limiting tidal radii of the clusters are examined by 
(1) the spatial surface density map of stars, 
(2) the radial profiles of the surface density, and 
(3) the azimuthal number density profiles as a function of
position angle at various radial distances from the cluster center.

To construct a spatial surface density map, we first made a two-dimensional 
star count map by binning into small grids on the spatial stellar distributions of the
C-M mask-selected stars for each cluster.
The background density level was simply assigned by the measured mean 
surface density value for the region outside the tidal radius of each cluster
since a variation with relatively low background density does not seriously affect
to the two-dimensional configuration of stellar distribution around a cluster.
The background-subtracted star count map was transformed into a 
smoothed surface density map by means of a Gaussian smoothing algorithm 
that enhanced the low spatial frequency of the background variation
and removed the high frequency of the spatial stellar density variation.
Then, we overlaid isodensity contour levels with a standard deviation unit ($\sigma$)
of the background level on the smoothed maps with various kernel values.

As an analytic indicator to confirm the overdensity feature around the tidal radius 
of a cluster, the radial surface density profile of a cluster was derived 
by measuring the number of stars in concentric annuli with a $1.0^{'}$ width.
The effective radius of each annulus is assigned by the equation $r_{e}=\sqrt{0.5(r_i^2+r_{i+1}^2)}$, where
$r_i$ and $r_{i+1}$ are the inner and outer radii of an annulus, respectively.
Radial surface density profiles for a different direction are also derived, for which we
divided an annulus with $3.0^{'}$ widths into eight sections (S1$-$S8) with an 
angle of $45^{\circ}$, as shown in Figure~\ref{annuly}. 
Here, we caution that the radial surface densities at the central region of cluster fields
are usually underestimated by crowding effects.
In order to examine the crowding effects, 
we added appropriate number of artificial stars with $i$ magnitudes in 
the range of the C-M mask-selected area (Figure ~\ref{cmd1}) to the $i$ images of each cluster, and 
we calculated the mean radial recovery rates of the input artificial stars
as completenesses in photometric measurements
out to $\sim25^{'}$ from the cluster's center.
Considering the determined photometric completenesses,
we empirically fit the ~\citet{King66} models to 
the observed radial surface density profiles of target clusters
over a radial range that keeps up recovery rates in stable with high completenesses.
It is worth noting that 
~\citet{Mcl05} have used the ~\citet{Wil75} models 
to describe properly the outer surface density structure 
of the majority of globular clusters in the Galaxy and the Magellanic Clouds. 
It is true that the Wilson models are spatially extended more than the ~\citet{King66} models,
but the Wilson models have still finite truncated tidal radii. 
This leads that the Wilson models are unlikely to describe extratidal overdensity features of globular clusters,
which are usually characterized by a break and radially untruncated power laws 
of the surface density profiles.

Clusters with obvious extratidal extensions 
showed a break in the observed radial surface density profiles near the tidal radius
~\citep{Gri95,Leo00,Tes00,Roc02,Lee03,Ols09}, which departs from the form predicted by the~\citet{King66} models.
Numerical simulations also confirmed such a break with power laws
in the radial surface density profiles due to the features of 
tidal debris around the clusters~\citep[e.g.,][]{Com99,Joh99,Joh02}.
Indeed, ~\citet{Joh99} predicted that for a constant mass-loss rate,
the radial surface densities of stripped stars in the tidal tails are proportional to $r^{-1}$.
However, ~\citet{Roc02} found that the annular-averaged density of stars along 
the well-defined extratidal tails around Pal 5 
is fit to a power law with a significantly steeper slope than that predicted by ~\citet{Joh99}.
For the target clusters in this paper, 
we fit a power law of $r^\gamma$ to the outer part of radial profiles measured in 
full concentric annuli, and compare the measured slope ($\gamma$) with those of ~\citet{Roc02} and ~\citet{Joh99}.
To investigate directions of the overdensity feature,
we also measure slopes of power laws for 
the radial surface density profiles to eight angular sections with different directions.

The azimuthal number density of stars with respect to a position angle 
was also derived for stars in an assigned annulus.
The position angle was measured clockwise from the east principal axis in a 
$10^{\circ}$ unit. 
Widths of each annulus were assigned to be $10^{'}$ and $15^{'}$ 
for the inner and outer regions, respectively, in order to contain a statistically 
sufficient number of stars in each area.
The radially cumulative azimuthal number density profile was also derived 
to trace the direction of the stellar overdensity feature around the clusters.

\subsection{M15}

Figure~\ref{M15con} shows a raw star count map around M15 and 
the surface density maps smoothed with various Gaussian kernel values
of $0.09^{\circ}, 0.135^{\circ}$, and $0.195^{\circ}$, from the upper-left panel 
to the lower-right panel.
The spatial bin size of the star count map was originally set to be $1.8^{'}\times1.8^{'}$.
Several individual blank chips, which did not contain useful observational data
in Figure~\ref{M15con}, were excluded from the subsequent analyses.
Isodensity plots were overlaid on the maps with the contour levels of 
$2\sigma, 2.5\sigma, 3\sigma, 4\sigma, 8\sigma$, and $30\sigma$ above 
background density level, 
where $\sigma$ is the standard deviation of the mean background value.
Contours in the star count map are the same as those in the
smoothed surface density map with a Gaussian kernel value of $0.195^{\circ} $.
The different arrows indicate directions of the Galactic center (short arrow) and 
the projected absolute proper motion (long arrow), i.e.,
$\mu_{\alpha}\cos\delta=-0.30\pm1.00$ mas yr$^{-1}$ and $\mu_{\delta}=-4.20\pm1.00$ mas
yr$^{-1}$ ~\citep{Cud93}.
The circle centered on the cluster indicates its tidal radius of $r_t=21.5'$ 
~\citep{Har96}.

It is apparent in Figure~\ref{M15con} that 
the overdensity feature is obviously extended over the tidal radius of M15 
at the levels larger than $2\sigma$ above the background.
Furthermore, the extratidal overdensity feature beyond the tidal radius
seem to form extended tails around the cluster. Among them, 
long tails to the southwest and northeast directions around the cluster 
are aligned apparently with the directions of the Galactic center and anticenter.
Marginal extensions are present in the outskirts to the south and north directions, 
which could correspond to the cluster's orbit.
The overdensity feature of the tails becomes more prominent in the lower resolution 
surface density maps with a larger Gaussian kernal value.
It is noted here that we did not find distinctive evidence 
for a southeast tidal extension as observed by ~\citet{Gri95}. 
The difference could be accounted for by the lower spatial resolution
and the brighter photometric limiting magnitude used by them. 
Indeed, most of the cluster stars in this study are faint main-sequence stars,
while \citet{Gri95} used the subgiant branch stars and main-sequence stars near 
the turn-off point of the CMD.
Note that ~\citet{Com99} certified in their simulation that
low mass stars were usually stripped from a globular cluster and formed
tidal tails or streams. 

\begin{figure}
\centering
\includegraphics[width=0.7\textwidth]{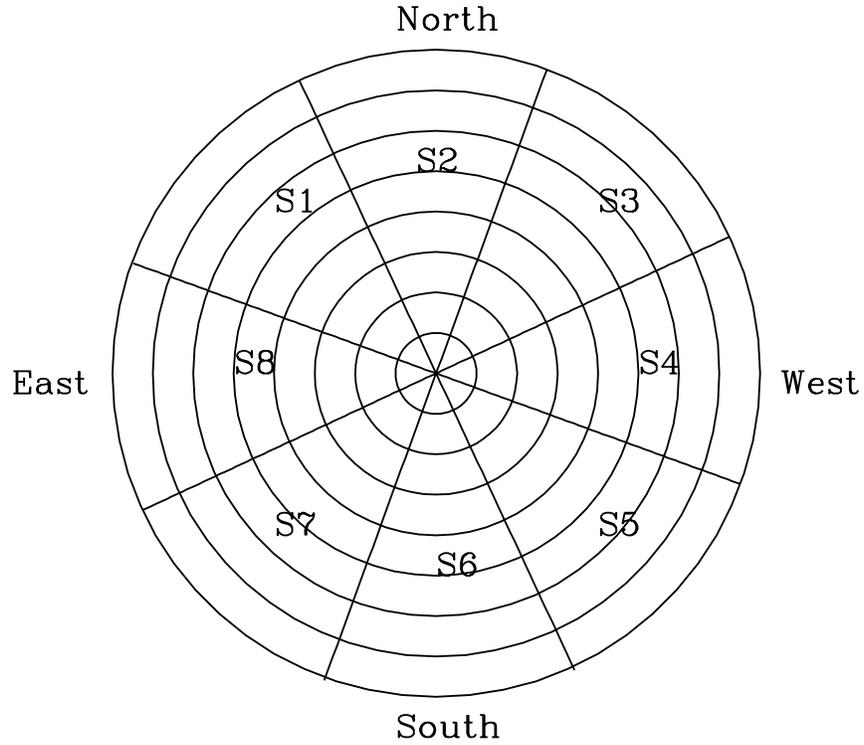}
\caption{
The schematic reseau plot used for the star counts. 
The radial surface densities were measured in concentric annuli. 
Each annunus was divided into eight angular sections (S1$\sim$S8)
to derive surface density profiles for a different direction. 
Azimuthal number densities were determined in an assigned annulus 
with respect to a position angle measured clockwise from the east principal axis
with a $10^{\circ}$ unit.
\label{annuly}}
\end{figure}

\begin{figure}
\centering
\includegraphics[width=0.7\textwidth]{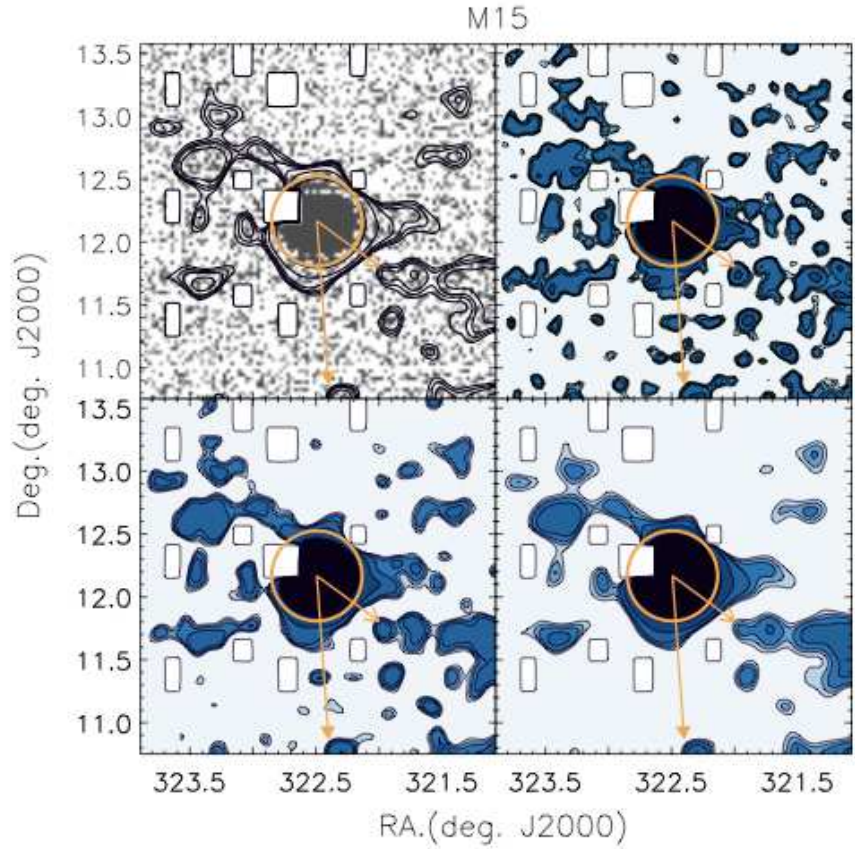}
\caption{From top-left to bottom-right, 
the raw star count map around M15 and the surface density maps
smoothed by Gaussian kernel values of $0.090^{\circ}$, $0.135^{\circ}$, and $0.195^{\circ}$,
overlaid with isodensity contour levels of $2\sigma, 2.5\sigma, 3\sigma, 4\sigma, 8\sigma$, 
and $30\sigma$ above the background level.
The circle indicates a tidal radius of M15.
The short and long arrows represent the direction of the Galactic center
and of the cluster's proper motion, respectively.
\label{M15con}}
\end{figure}

\begin{figure}
\centering
\includegraphics[width=0.3\textwidth]{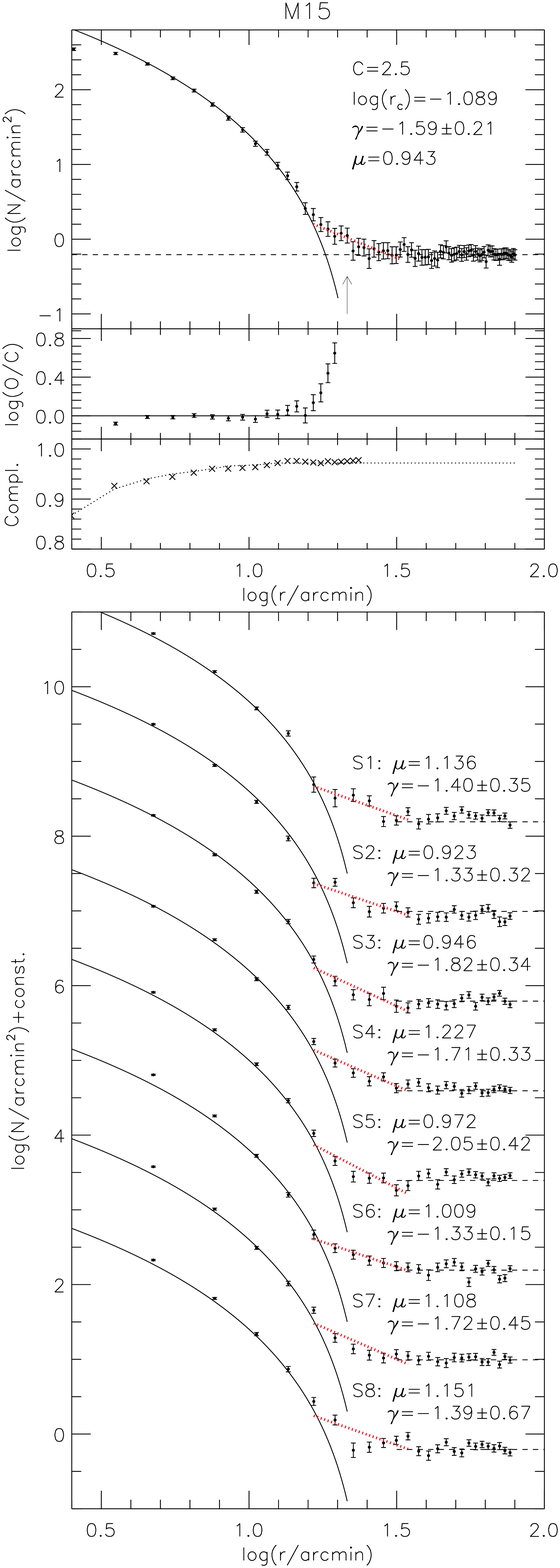}
\caption {{\it Upper} : Radial surface density profile of M15 with a theoretical King 
model of $c=2.5$ (solid curve). Residuals, log(O/C), after subtracting the values 
predicted by the King model from the observed surface densities are also plotted.
Radial completeness, the ratio of the number of recovered artificial stars 
and the total number of stars, and its trend (dotted line) are added.
The profile within the detected overdensity region, $\sim 0.7 r_t\leq r\leq \sim 1.5 r_t$, 
is represented by a power-law, i.e., a dotted straight line in logarithmic scale
with a slope of $\gamma=-1.59\pm0.21$.
The mean number density of stars in the overdensity region is estimated
as $\mu=0.943$ per arcminute square. The dashed horizontal line indicates the 
background density level, and an arrow represents the tidal radius of M15.
{\it Lower} : Radial surface density profiles of eight different angular 
sections (S1$\sim$S8). For clarity, the surface densities from S1 to S8 
are given zero-point offsets with values of const.= 8.4, 7.2, 6.0, 4.8, 3.6, 2.4, 1.2, 0.0.
Dotted lines indicate the power-law fit to the overdensity region of 
$\sim 0.7 r_t\leq r\leq \sim 1.5 r_t$ with $\gamma$ as a slope.
$\mu$ is the mean density in the overdensity region.\label{M15king}
}
\end{figure}

The radial surface densities for M15 measured in each concentric annulus are plotted 
in the upper panel of Figure ~\ref{M15king} along with a theoretical 
King model of $c=2.5$, which are arbitrarily normalized to our measurements. 
We also show two plots of the observed-minus-predicted residuals, $log(O/C)$, 
to illustrate the departures of the observed radial surface densities 
from the prediction of the King model,
and of the photometric recovery rates to consider radial crowding effects
on the observed profile.
It is apparent in the plots that incompleteness due to crowding effects
reduces the measured surface densities in the innermost region, while
the radial surface density profile in the region outer than $log(r^{'})\sim0.7$ 
has completeness values larger than $\sim 95\%$.
Consequently, the radial surface density profile departs from the King model
with a break at a radius less than the tidal radius, i.e., $\sim0.7r_t$, and 
the excess extends to $\sim1.5r_t$.
The radial surface density profile in this overdensity region 
resembles a radial power law with a slope of $\gamma =-1.59\pm0.21$,
which is comparable to the value  $-1.58\pm0.07$ proposed by ~\citet{Roc02}
from the annular-averaged densities of stars along the extratidal tails of Pal 5.
However, the slope is significantly steeper than the case of $\gamma = -1$, predicted 
for a constant mass-loss rate over a long time ~\citep{Joh99}.
Radial surface density profiles to eight angular sections with a different direction,
as illustrated in Figure~\ref{annuly}, are also represented in the lower panel of 
Figure ~\ref{M15king}.
The overdensity feature in the region of $0.7r_t\lesssim r \lesssim 1.5r_t$ 
is commonly detected in all eight angular sections.
The estimated mean surface densities ($\mu$) in the angular sections 1, 4 and 8
are higher than those in the other sections. This is in good agreement 
with the extended overdensity features to the directions of the Galactic 
center and anticenter, as shown in the surface density maps of Figure~\ref{M15con}.
Comparing the power laws of the radial surface density profile in each angular section,
the slopes ($\gamma$) seem to be shallower in angular sections 2 and 6, 
which are likely aligned with the orbit of M15, i.e., the direction of the cluster's 
proper motion.

\begin{figure}
\centering
\includegraphics[width=0.7\textwidth]{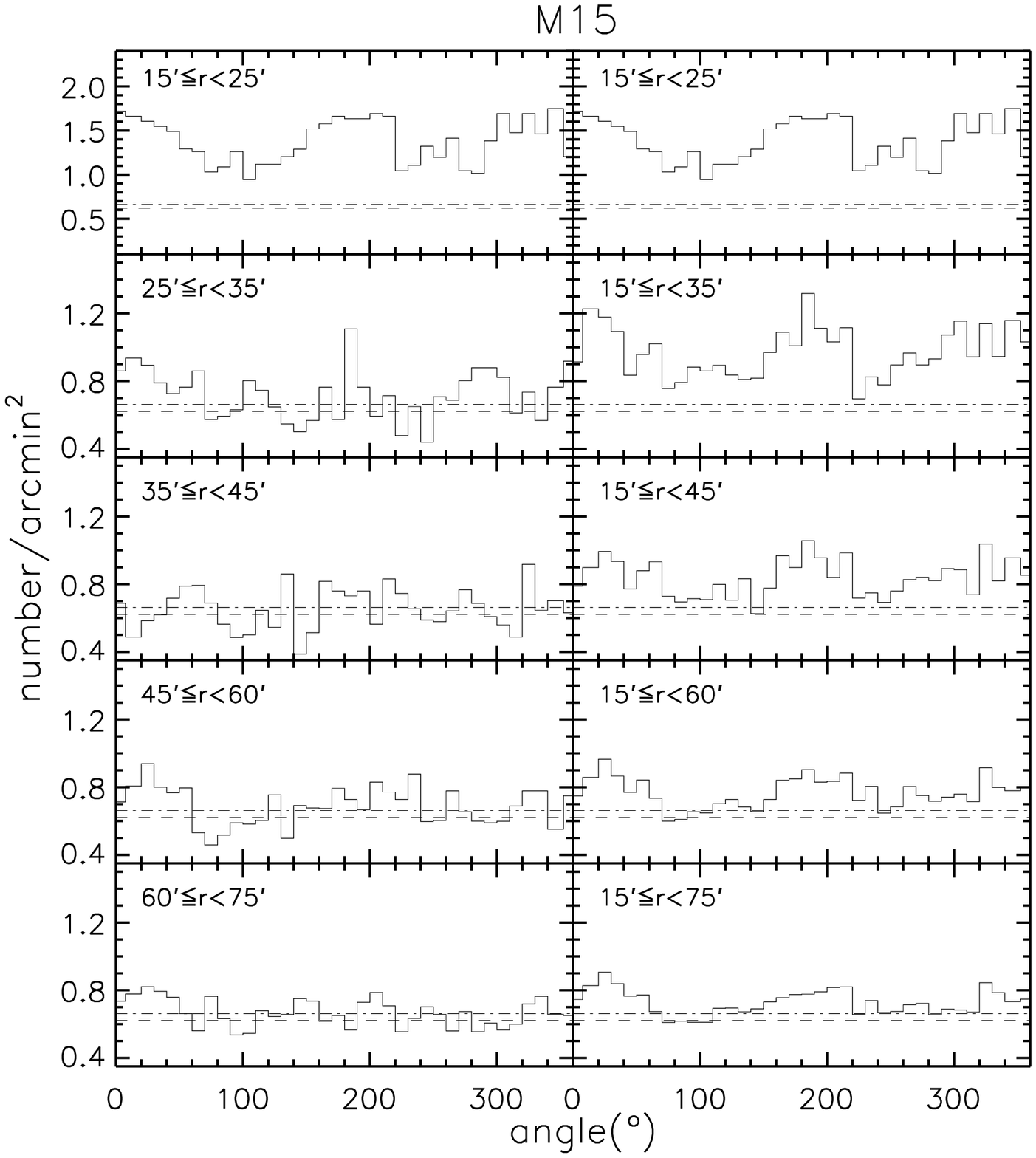}
\caption{The annular (left panels) and radially cumulative (right panels)
azimuthal number density profiles of M15 with an angular bin size of $10^{\circ}$.
The radial bins are set to be $10^{'}$ for the inner three annuli and $15^{'}$ for the 
outer two annuli. 
The horizontal dashed lines are the determined backgound density level and 
dashed dotted lines are the level of 1$\sigma$ deviation above background level.
\label{M15azim}}
\end{figure}

Figure ~\ref{M15azim} shows the annular and cumulative azimuthal number density profiles
around M15. 
As presented in Figure ~\ref{annuly}, we measured the number density clockwise 
from the east principal axis with a $10^{\circ}$ angular bin.
Note that number densities at the angular bins without data, caused by the blank chips, 
are interpolated by the values of adjacent angular bins.
The radial range $15^{'}\leqq r < 75^{'}$  corresponds to $0.7r_t\lesssim r \lesssim 3.5r_t$,
in which we examine the direction and the extension of the tail feature of M15.
The radial bins are set to be $10^{'}$ for the inner three annuli and $15^{'}$ for the 
outer two annuli.
At the innermost annulus with $15^{'}\leqq r < 25^{'}$, 
it is apparent that the number densities in the ranges of position angles 
$\sim150^{\circ}$ to $\sim220^{\circ}$ and $\sim300^{\circ}$ to $\sim60^{\circ}$ 
are clearly higher than those in the other position angles.
The high density features in the positions are likely to extend 
to the whole observed field on the sky, as shown in the cumulative azimuthal number 
density profile.
This is in accordance with the tail features to the direction of the 
Galactic center and anticenter, as shown in the surface density map of Figure~\ref{M15con}. 
With a somewhat low confidence, there also seems to be relatively weak 
overdensity features at position angles 
$\sim100^{\circ}$ and $\sim280^{\circ}$ in the regions of the inner two annuli
(i.e., $15^{'}\leqq r < 35^{'}$).
These are reliably matched with the tails to the direction along the cluster's orbit, 
as inferred from the surface density maps and the radial density profiles of M15.

\subsection{M30}

\begin{figure}
\centering
\includegraphics[width=0.7\textwidth]{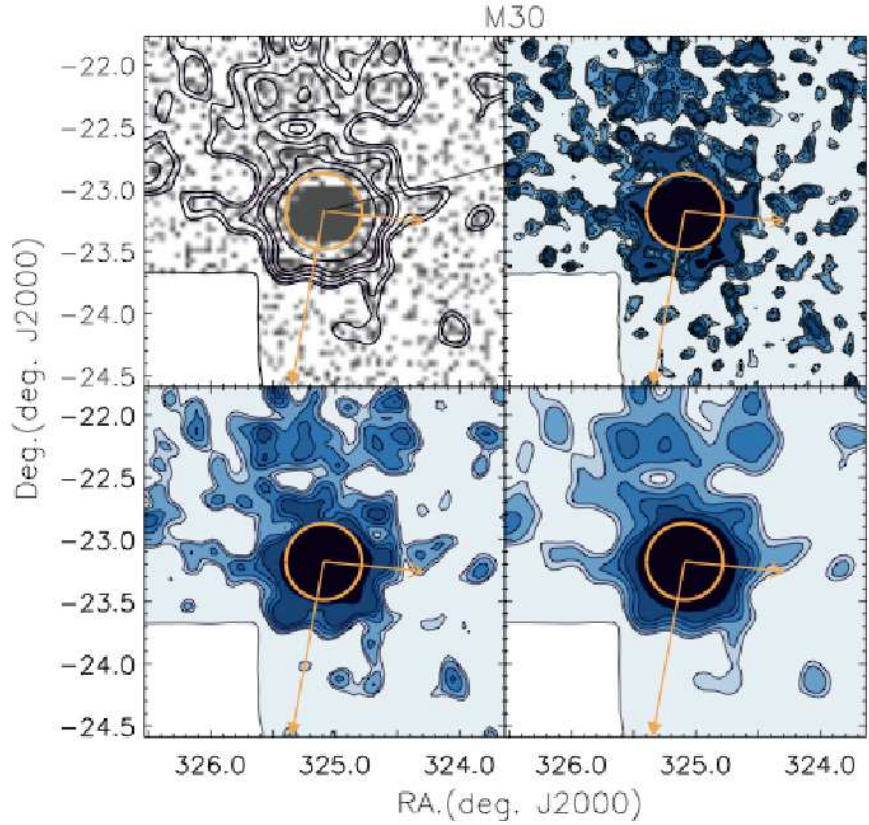}
\caption{The raw star count map around M30 and the surface density
contour maps smoothed by Gaussian kernel values of $0.084^{\circ}, 0.147^{\circ}$, 
and $0.210^{\circ}$. The contour levels are the background level, $0.5\sigma,1.5\sigma,
2.5\sigma,3.5\sigma$, and $8\sigma$ above the background level.
The circle indicates a tidal radius of M30.
The short and long arrows represent the direction of the Galactic center
and of the cluster's proper motion, respectively.\label{M30con}}
\end{figure}

In Figure \ref{M30con} we show the spatial configuration of stars surrounding M30.
The top-left panel shows the two-dimensional star count map that was made 
by measuring the number of cluster stars in each $2.5^{'}\times2.5^{'}$ bin 
on the C-M mask-selected star count map of M30.
The other panels to the bottom-right show the smoothed surface density maps 
with Gaussian kernel widths of $0.084^{\circ}, 0.147^{\circ}$, and $0.210^{\circ}$.
Note that in Figure \ref{M30con} the Megacam observations were not secured 
at the outer southeast field of M30.
The isodensity contours with the background level and $0.5\sigma,1.5\sigma, 2.5\sigma,
3.5\sigma$ and $8\sigma$ levels above the background were overlaid on the surface density maps.
Contour levels in the star count map correspond to those of a smoothed map with 
a Gaussian kernel width of $0.210^{\circ}$.
The short arrow indicates the projected direction to the Galactic center on the sky,
and the long arrow indicates the projected proper motion with
$\mu_{\alpha}\cos\delta=1.42\pm0.69$ mas yr$^{-1}$ and  $\mu_{\delta}=-7.71\pm0.65$
mas yr$^{-1}$, measured by~\citet{Din99}.
The tidal radius of M30 (i.e., $r_t=18.34'$) from~\citet{Har96} was
also plotted as a circle centered on the cluster.

At contour levels larger than $0.5\sigma$ above the background,
it is apparent in Figure \ref{M30con} that a flocculent 
overdensity feature with tails is well developed around M30
and extends to about $2r_t\approx40^{'}$ and beyond.
While there is no strong evidence of extratidal features extending 
into specific directions in the surface density maps,
the apparent east-west extension seems to be aligned 
with the direction of the Galactic center and anticenter.
There is no apparent extended extratidal tail toward the direction of the 
proper motion on the cluster's orbit in the surface density maps.
Instead, distant stars unbound from the cluster are likely 
building up a feature of the northern extension to the opposite direction 
of the cluster's proper motion, with significant overdensities up to a contour level of
$2.5\sigma$ in Figure \ref{M30con}.
The northern extension feature
is likely tracing the cluster's trailing tail through its orbital path.
We note, however, given the position and distance of M30 from the Galactic plane,
the northern extension of the stars could be a result of gravitational interaction
with the Galactic plane~\citep{Pio97}.

\begin{figure}
\centering
\includegraphics[width=0.3\textwidth]{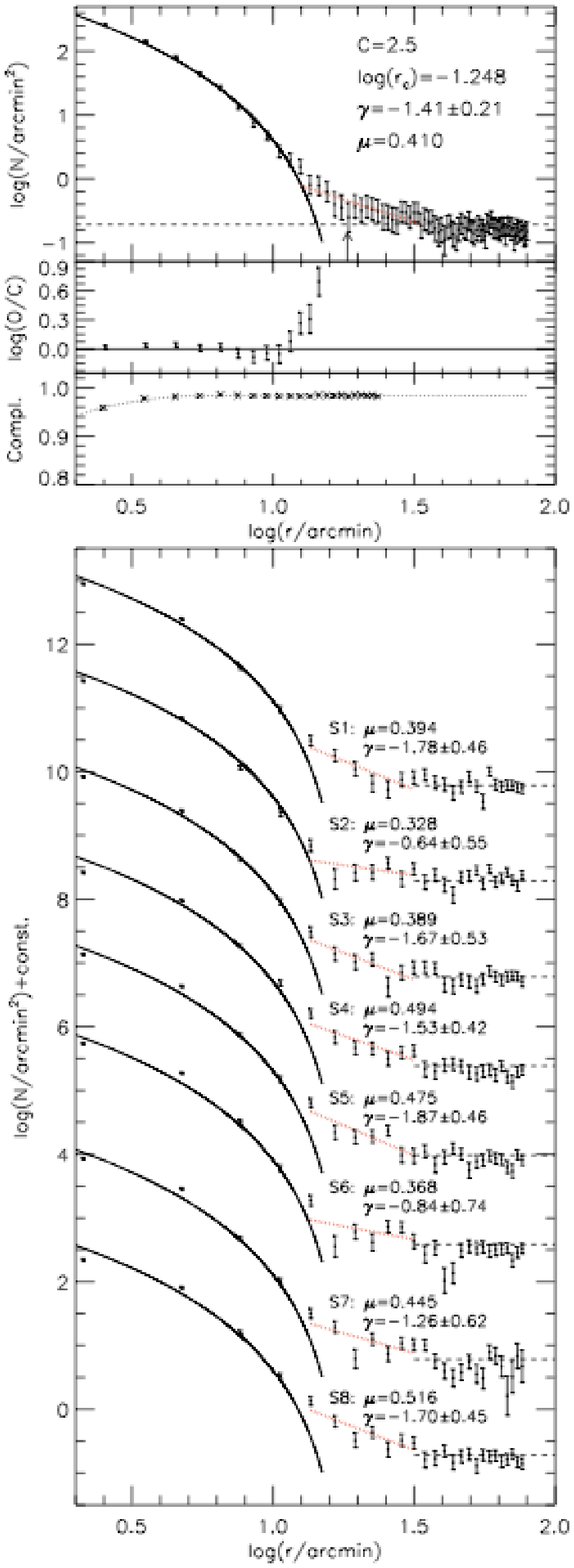}
\caption {{\it Upper} : Radial surface density profile of M30 with a theoretical King 
model of $c=2.5$ (solid curve), residuals after subtracting the values predicted by 
the King model, and radial completeness and its trend (dotted line).
The profile within the detected overdensity region, $\sim 0.7r_t\leq r\leq \sim 2r_t$, 
is represented by a power-law, i.e., a dotted straight line in logarithmic scale
with a slope of $\gamma=-1.41\pm0.21$.
The mean number density of stars in the overdensity region is estimated
as $\mu=0.410$ per arcminute square. 
The dashed horizontal line indicates the 
background density level, and an arrow represents the tidal radius of M30.
{\it Lower} : Radial surface density profiles to eight different angular 
sections (S1$\sim$S8). For clarity, the surface densities from S1 to S8 
are given zero-point offsets with values of 
const.= 10.5, 9.0, 7.5, 6.1, 4.7, 3.3, 1.5, 0.0.
Dotted lines indicate the power-law fit to the overdensity region with $\gamma$ as a slope.
$\mu$ is the mean density in the overdensity region.\label{M30king}
}
\end{figure}

A radial surface density profile for M30 was derived by measuring the mean number density
of stars in concentric annuli and is presented in the upper panel of Figure~\ref{M30king}.
For the inner profile, we overplotted a theoretical King profile of $c=2.5$, 
which was arbitrarily normalized to the observed radial surface density profile of M30. 
The observed-minus-predicted residuals illustrate 
the deviations of the observed radial surface densities from the prediction
of the King model. 
The innermost region shows crowding effects out to $log(r^{'})\sim0.5$,
after which the photometric completenesses keep a nearly constant value of $\sim 98\%$.
A break in the observed radial density profile appears at a radius less than 
the tidal radius of M30, i.e., $\sim0.6r_t$, and the overdensity feature extends to 
about $2r_t$ of the cluster.
The departure of the observed radial surface density profile in this region from the 
King profile is characterized by a power law with a slope of $\gamma=-1.41\pm0.21$.
Similar to the case of M15, the slope is also significantly steeper than the index of $r^{-1}$ predicted 
for extratidal structure with a constant mass-loss rate over a long time ~\citep{Joh99}.
Radial surface density profiles to eight angular sections are plotted in
the lower panel of Figure~\ref{M30king}.
Note that incompleteness due to crowding reduces the measured surface densities at the innermost point
with respect to those expected by the King model.
It is apparent that the profiles have in common
a break at $\sim0.6r_t$ and an extended overdensity feature to $\sim2r_t$ with a power law.
At the radial range, the estimated mean surface densities are relatively
higher in the angular sections 4 and 8 than in the other sections. This can be explained by
the spatially extended extratidal features to the directions of the Galactic center 
and anticenter, as shown in the surface density map of Figure~\ref{M30con}.
We also note that the slopes of the radial density profiles of the angular sections 2 and 6
are shallower in the radial range of $0.6r_t\lesssim r \lesssim 2r_t$ than those 
of the other angular sections. This feature may be related to a cluster's orbit inferred 
from the proper motion, although the spatial surface density maps do not show an 
extinctive extratidal feature of stars on the cluster's orbit.

\begin{figure}
\centering
\includegraphics[width=0.7\textwidth]{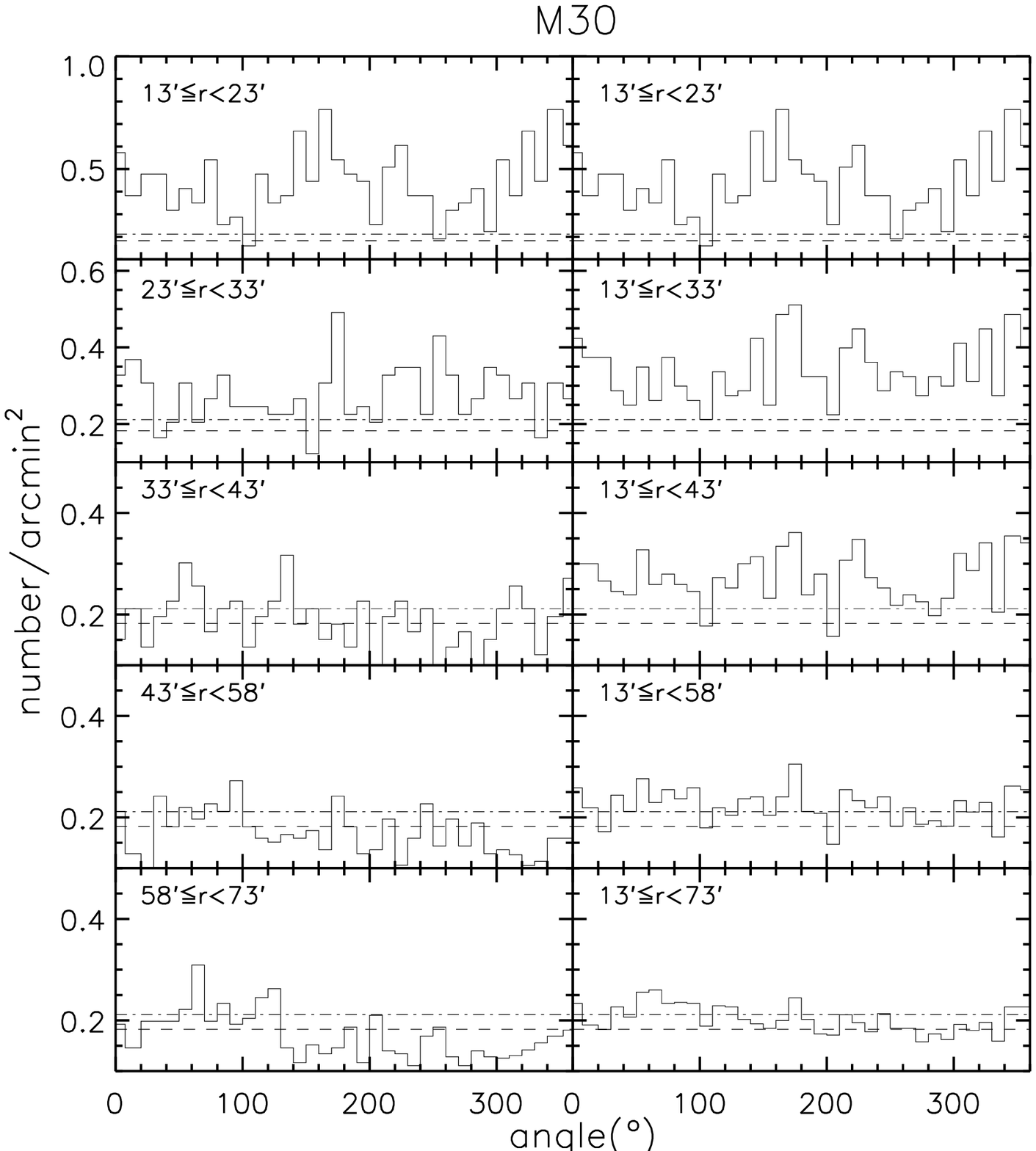}
\caption{The annular (left panels) and radially cumulative (right panels)
azimuthal number density profiles of M30 with an angular bin size of $10^{\circ}$.
The radial bins are set to be $10^{'}$ for the inner three annuli and $15^{'}$ for the 
outer two annuli. 
The horizontal dashed lines are the determined backgound density level and 
dashed dotted lines are the level of 1$\sigma$ deviation above background level.
\label{M30azim}}
\end{figure}

The left and right panels of Figure~\ref{M30azim} show 
the annular and radially cumulative azimuthal number density profiles, respectively, 
of M30 with an angular bin of $10^{\circ}$.
The radial bins are set to be $10^{'}$ for the inner three annuli and $15^{'}$ for the 
outer two annuli. The tidal radius of M30 is placed in the innermost annulus.
As shown in Figure~\ref{M30azim} the azimuthal number density profiles 
have a flocculent feature with respect to the position angle, which is also expected in 
the spatial surface density maps of Figure~\ref{M30con}.
Nevertheless, there appear to be two main overdensity ranges of the position angle, 
i.e., $\sim120^{\circ}$ to $\sim220^{\circ}$ and $\sim300^{\circ}$ to $\sim40^{\circ}$,
in the azimuthal number density profile for the innermost annulus.
Also, the overdensity feature in the ranges of the position angle extends to 
the second annulus, i.e. $\lesssim2r_t$, with a somewhat low confidence.
This might be associated with the apparent east-west extension of stars 
to the direction of the Galactic center and anticenter, as shown in the spatial 
surface density maps in Figure \ref{M30con}.
In the azimuthal number density profiles, we did not find evidence of 
the extended extratidal features induced by the cluster's orbital motion 
at the remote area in the vicinity of M30.

\subsection{M53 and NGC 5053}

Figure ~\ref{N5053M53contour} represents the spatial configuration of selected stars 
in the vicinity of a pair of the two clusters M53 and NGC 5053. 
As described in Sections 2 and 3, cluster stars for both clusters are selected 
by using the same selection criteria in C-M spaces of $(c_1, i^{'})$ 
and  $(c_2, i^{'})$. The upper-left panel is a raw star count map with a 
spatial bin size of $3.0^{'}\times3.0^{'}$.
The surface density maps smoothed with various Gaussian kernel values 
of $0.125^{\circ}, 0.175^{\circ}$, and $0.240^{\circ}$ are also plotted in 
Figure ~\ref{N5053M53contour} to bottom-right.
The long arrow indicates the direction of the proper motion of M53, and the
two short arrows represent the direction to the Galactic center. 
The proper motions of M53, i.e., $\mu_{\alpha}\cos\delta=0.50\pm1.00$ mas yr
$^{-1}$, $\mu_{\delta}=-0.10\pm1.00$ mas yr$^{-1}$ are adopted from
~\citet{Din99}. The proper motion of NGC 5053 has not yet been reported.
Circles are tidal radii for each cluster, i.e., $r_t=21.75^{'}$ for M53 and 
$r_t=13.67^{'}$ for NGC 5053~\citep{Har96}.

It is remarkable in Figure ~\ref{N5053M53contour} that the spatial distribution of 
stars in the vicinity of M53 and NGC 5053 show a complex configuration 
with various extratidal features. The contour levels are
background level, and $2\sigma,4\sigma,8\sigma$, and $20\sigma$
above background level, where $\sigma$ is the standard deviation of the estimated 
background level. The main new findings that were observed in the surface density 
maps are as follows: 
(1) the extratidal overdensity features extended over the tidal radii of both clusters
at the contour levels above the background, 
(2) features of tails on the cluster's orbit
in the east-west direction for M53 and to the direction of the Galactic center for NGC 5053, 
(3) complex substructures of extratidal feature of clumps and ripples around the clusters, 
(4) a tidal bridge-like structure connecting spatially 
the close neighbor pairing of M53 and NGC 5053, and finally 
(5) an envelope composed by stars surrounding the cluster system.
Of particular interest are the tidal features that are possibly enhanced by the dynamical 
interaction between two clusters,
which include features such as clumps and ripples surrounding the clusters and most 
strikingly, a tidal bridge-like feature and an envelope structure around the multiple 
globular cluster system. 
Note that there are no known binary globular clusters in the Milky Way,
while ~\citet{Min04} reported the discovery of a binary cluster in the peculiar giant 
elliptical galaxy NGC 5128.
We therefore point out that these observed extratidal structures 
of the pairing system of M53 and NGC 5053 could
indeed play a unique key role in the effect of near passage 
on the dynamical tidal interaction between two globular clusters in the Galaxy.

The upper panel of Figure~\ref{M53king} shows the radial surface density profile of M53, which
was derived by measuring mean stellar number densities in each concentric annulus.
Stars within the tidal radius of the companion cluster NGC 5053 were not included 
in the radial surface density profile of M53
to avoid overlapping of stars in NGC 5053.
In order to compare the observed radial density profile,
a theoretical King model was computed with a concentration parameter $c=1.78$.
As apparent in Figure~\ref{M53king}, the observed profile departs from the King model
with a break at a radius less than the tidal radius, i.e., $0.7r_t$.
Note that the surface densities in the innermost region is affected by crowding effects 
out to $log(r^{'})\sim0.8$, after which
the photometric completenesses show a nearly constant value of $\sim 98\%$,
as shown in the plot of the radial photometric recovery rates.
The overdensity feature, which extends to $1.6r_t$, is likely represented by a power law 
with a slope of $\gamma=-1.58\pm0.19$, 
which is almost identical with the value $-1.58\pm0.07$ for the annular-averaged densities of
stars in the extratidal tails of Pal 5 ~\citep{Roc02}, but
significantly steeper than that predicted from a constant orbit-averaged mass loss rate \citep{Joh99}.
The residual plot within the tidal radius
also exhibits a departure of the observed radial surface densities 
from the prediction of the theoretical King model.
As a signature of the presence of tidally stripped stars,
~\citet{Bec08} also recognized a change in slope of the 
radial surface density profile for M53.

The lower panel of Figure~\ref{M53king} shows radial surface density 
profiles of M53 in different directions. Of the total eight angular sections defined 
in Figure~\ref{annuly}, 
we could not derive the radial profiles of the angular sections 5 and 6 
because observational data were not secured for the southern area of M53.
As can be seen from Figure~\ref{M53king}, the obtained radial profiles for the 
six angular sections show
behaviors of a break at $\sim0.7r_t$ in common, becoming power laws over the tidal 
radius to $1.6r_t$. This trend resembles the extended overdensity features 
over the tidal radius appearing in the surface density maps of Figure~\ref{N5053M53contour}.
The estimated mean densities ($\mu$) at the radial range show higher 
values in the angular sections 1, 4, and 8.
This feature corresponds to the apparent extratidal tails on the cluster's orbit
in the east-west direction aound M53, as shown in Figure~\ref{N5053M53contour}.
Also, slopes ($\gamma$) of the power laws in the angular sections 3 and 4 seem 
to be slightly flatter 
than the mean slope of the radial surface density profile of M53.
It can be noted that angular sections 3 and 4 are equivalent in the directions 
of the Galactic anticenter and the trailing tail on the cluster's orbit.

In Figure~\ref{M53azim}, we show the annular and cumulative azimuthal 
number density profiles around M53 with a $10^{\circ}$ angular bin. 
We have divided the total sample in the radial range of $15^{'}\leqq r \leqq 75^{'}$,
which corresponds to $0.7r_t\lesssim r \lesssim 3.5r_t$,
into five concentric annuli with radial bins with
$10^{'}$ and $15^{'}$ widths for the inner three and outer two bins, respectively.
The tidal radius of M53 is placed in the innermost annulus.
The blank bins in Figure~\ref{M53azim} correspond to 
the areas without photometric data and 
the area within the tidal radius of the neighbor cluster NGC 5053.
The azimuthal density profiles for the two inner annuli indicate 
the overdensity feature in the range of position angles $\sim160^{\circ}$ to 
$\sim240^{\circ}$, which is in good agreement with the direction of the trailing 
tail on the cluster's orbit. Also, an apparent overdensity feature in the 
range of position angles $\sim310^{\circ}$ to $\sim20^{\circ}$
likely traces the tidal bridge-like feature in the direction of the companion 
cluster NGC 5053, as shown in the surface density maps of Figure~\ref{N5053M53contour}.

We now turn to NGC 5053, which is a very diffuse cluster
with a low concentration parameter $c=0.85$ ~\citep{Har96}.
Figure~\ref{N5053king} shows the derived radial surface density profiles of NGC 5053.
As in the case of the companion cluster M53, stars within the tidal radius of M53 were 
not used to derive the radial density profile of NGC 5053.
The upper panel shows the profile of the mean surface densities measured in 
each concentric annulus, and the lower panel consists of those of eight angular sections.
The observed profiles were compared with a theoretical King model of $c=0.85$.
By crowding effects, surface densities seem to be reduced in the 
central region of the cluster inner than $log(r^{'})\sim0.5$, after which
the photometric completenesses show a nearly constant value of $\sim 96\%$.
A break at $\sim0.7r_t$ is commonly appeared in the obtained radial density profiles.
The overdensity feature extended to $2.5r_t$ of NGC 5053.
For the radial profiles measured in concentric annulii,
the departure from the King model in this region
can be represented by a power law with a slope $\gamma=-0.62\pm0.15$.
It is remarkable that the radial surface density profiles of NGC 5053 are characterized by 
a long extension of the extratidal overdensity feature and a much shallower slope
of the power law compared with the case of the slope $\gamma=-1$ in a constant mass-loss
rate of ~\citet{Joh99}. This is indeed consistent with the diffuse structure of the cluster
with a small concentration of $c=0.84$ ~\citep{Har96}.
In the overdensity region, the mean surface densities ($\mu$) are higher
in the angular sections 3 and 4, and the angular sections 7 and 8
than in the other angular sections. Angular sections 3 and 4 are coincident with 
the tidal bridge-like feature toward the neighbor cluster M53. 
Also, the high density in angular sections 7 and 8 might be related to the apparently extended 
extratidal features in the direction of the Galactic center, as shown in 
Figure~\ref{N5053M53contour}.

Figure~\ref{N5053azim} shows the annular and radially cumulative azimuthal number 
density profiles around NGC 5053, as a function of position angle.
The radial range $10^{'}\leqq r \leqq 80^{'}$ corresponds to 
$0.7r_t \lesssim r \lesssim 5.9r_t$,
in which we set six concentric annuli.
The blank bins correspond to the areas without photometric data and 
the area within the tidal radius of the neighbor cluster M53.
It is apparent that the azimuthal number density profile shows a flocculent 
feature at the innermost annulus in which the tidal radius of the cluster is imbedded.
Peaks in the range of position angle $\sim100^{\circ}$ to $\sim200^{\circ}$ represent
the tidal bridge-like feature appearing in the surface density maps.
An extratidal extension feature toward the Galactic center is correlated with 
peaks in the range of $\sim250^{\circ}$ to $\sim50^{\circ}$ on the azimuthal number 
density profiles.

\subsection{NGC 5466}

To examine the spatial configuration of stars around NGC 5466, 
we first constructed a star count map,
as shown in the top-left panel of Figure~\ref{N5466con},
with a spatial bin size of $2.88^{'}\times2.88^{'}$ for $\sim3^{\circ}\times3^{\circ}$ 
total field of view. We then smoothed the star count map with Gaussian kernel widths of 
$0.168^{\circ}, 0.264^{\circ}$, and $0.312^{\circ}$,
yielding smoothed surface density maps shown in Figure~\ref{N5466con}.
Isodensity contour levels in Figure~\ref{N5466con} correspond to values of the 
background level and $1\sigma, 2\sigma, 3\sigma, 4\sigma, 6\sigma, 10\sigma$, 
and $40\sigma$ above the background.
Contour levels in the star count map correspond to those of a smoothed map with
a Gaussian kernel width of $0.312^{\circ}$. 
The circle centered on the cluster indicates its tidal radius of 
$r_t=20.98^{'}$~\citep{Leh97}.

As can be seen in Figure~\ref{N5466con}, an overdensity feature above the 
background level apparently extends beyond the tidal radius of NGC 5466. 
Moreover, the extending extratidal overdensity feature seems to form two tails
in the southeast and in the northwest directions. 
The tails stretch out symmetrically to both sides of the cluster, 
tracing over $\sim2^{\circ}$ of arc in the observed field.
These are the same features reported by ~\citet{Bel06a} and ~\citet{Gril06}
in which the orientation of the tails is consistent with the cluster's orbit,
as judged from the proper motion data. Note that the long arrow in  Figure~\ref{N5466con}
indicates the direction of the projected proper motion of NGC 5466, 
i.e., $\mu_{\alpha}\cos\delta=-3.90\pm1.00$ mas yr$^{-1}$, 
$\mu_{\delta}=1.00\pm1.00$ mas yr$^{-1}$ ~\citep{Ode97},
while the short arrow represents the direction of the Galactic center.
The tails also weakly show an S-shape feature of tidal stripping for cluster stars,
as appeared for Palomar 5 ~\citep{Ode03}.
There are clumpy overdensity features in the southern area of the observed field,
which were also found in the surface density map of ~\citet{Gril06}.
However, neither the cluster's orbit nor the direction to the Galactic center is 
likely to have a correlation with the clumpy structure.
The appearance of such clumps can be caused either by the background galaxies
or by the local substructures of the stars in the Galactic halo.

The upper panel of Figure~\ref{N5466king} shows the radial surface density 
profile of NGC 5466, for which we measured mean stellar number densities 
in concentric annuli. The solid curve characterizes the empirical representation 
of the data fitted by the King model of $c=1.2$.
While the measured surface densities have a nearly constant completeness value of $\sim 98\%$,
the very central region of the cluster inner than $log(r^{'})\sim0.1$
seems to show crowding effects.
Although the inner part of the profile is well represented by the King model, 
the observed profile departs from the model prediction at the radius of $\sim0.5r_t$, 
showing the extended overdensity feature to be $\sim2r_t$. 
The radial surface density profile in this region is 
characterized by a power law with a slope $\gamma=-2.44\pm0.13$. 
The measured slope is significantly steeper than the cases
of a constant mass-loss rate $\gamma=-1$ ~\citep{Joh99} and of the other target clusters in this paper.
This is certainly because NGC 5466 shows a well-developed structure of extratidal tails.
Indeed, ~\citet{Roc02} found the annular-averaged surface densities of stars
along the well-defined extratidal tails of Pal 5 are also fitted to a power law in radius 
with a significantly steeper slope.
~\citet{Leh97} also found an increased surface density in the outer region of 
the radial density profile for NCG 5466, which may be an indication of a tidal tail
caused by unbound stars around the cluster.
The departure of the radial profile from the prediction of the King model 
is more obviously presented in the residual plot of Figure~\ref{N5466king}.  
The lower panel of Figure~\ref{N5466king} shows the radial surface density profiles 
of eight angular sections in the observed field, as mentioned in Figure~\ref{annuly}.
The break at $\sim0.5r_t$ and the increased density features out to $\sim2r_t$ 
are shown in profiles for all eight angular sections,
and are in good agreement with the extratidal overdensity feature in 
the surface density maps of Figure~\ref{N5466con}.
Also, the estimated mean surface densities ($\mu$) of the radial range are relatively 
higher in angular sections 3, 4, and sections 7, 8 than in the other sections. 
This is clearly a consequence of the two tails in the southeast and in the northwest 
directions around the cluster, as shown in the Figure~\ref{N5466con}.

The orientation of the detected two tails is further analyzed in Figure~\ref{N5466azim}, 
which shows annular (left panels) and radially cumulative (right panels) azimuthal 
number density profiles around NGC 5466.
The area within the radial range of $15^{'}\leqq r<75^{'}$, 
i.e., $0.7r_t\lesssim r \lesssim3.6r_t$ of NGC 5466,
was divided into five concentric annuli with radial bins of $10^{'}$ (inner three) 
and $15^{'}$ (outer two) widths.
We then clockwise measured the azimuthal number density with a $10^{\circ}$ angular bin.
It is apparent from the inner two annuli of the Figure~\ref{N5466azim} 
that there are two main overdensity ranges at the position angles
$\sim120^{\circ}$ to $\sim210^{\circ}$ and $\sim320^{\circ}$ to $\sim30^{\circ}$,
with peaks near $175^{\circ}$ and $345^{\circ}$, respectively.
The peaks seem to be slightly shifted to position angles $\sim140^{\circ}$ 
and $\sim330^{\circ}$ of the outer three annuli.
These are well matched with the symmetric S-shape feature of the extratidal tails 
detected in the surface density maps of stars around NGC 5466,
as shown in Figure~\ref{N5466con}.

\section{Discussions and Summary}

Using homogeneous wide-field observations obtained with the Megacam mosaic imager 
on the 3.6 m CFHT, we have undertaken several analyses to identify 
spatial configurations of stars surrounding five metal-poor globular clusters
in the Galactic halo, i.e., M15, M30, M53, NGC 5053, and NGC 5466.
Because of the low metallicities of the clusters we expected to easily detect 
the extratidal features from background stars 
with the secured wide-field deep photometric data 
covering an approximate $3^{\circ}\times3^{\circ}$ field-of-view.
Analyses of surface density maps revealed that all the clusters
present features of extratidal overdensities over their tidal radii
in the form of extended tidal tails and halos around the clusters.
The observed extratidal extensions likely trace the spatial orbits of the clusters
as well as the effects of dynamical interaction with the Galaxy.
Radial surface density profiles and azimuthal number density profiles
confirmed the overdensity feature and the extratidal structure of tails and halos
around the five metal-poor globular clusters.

M15, which is placed at $10.4$ kpc from the Galactic center, is one of the densest
globular clusters in the Milky Way. It is a post core-collapsed cluster ~\citep{deM96}
with a high concentration $c=2.5$ ~\citep{Har96}. 
~\citet{Lau06} pointed out that M15 is unlikely 
to have extratidal structures due to its high concentration.
However, we found apparent extratidal features from the analyses of the spatial stellar distribution 
around M15.
Indeed, surface density maps revealed features of tails in the directions of 
the Galactic center and anticenter and of tails in the direction of the cluster's orbit
inferred from the proper motion. 
The radial surface density profiles of stars surrounding M15 have a break at $\sim0.7r_t$
and an excess of surface density extending to $\sim1.5r_t$,
which indicate tidal tail features around M15.
~\citet{Gri95} and ~\citet{Leh97} also found an indication
for tidal tails in the radial stellar density profiles of M15. 
Radial surface density profiles in different directions 
and azimuthal number density profiles confirmed directions of 
the extended features of the tails around M15,
which are seen in the surface density maps. 
From the results, we can conclude that the apparent tails around M15
are real tidal features of the cluster.
No Abell clusters were found in the observed field, and 
the observed field was not polluted by a strong dust extinction (E$(B-V)=0.10$).
We note, however, that clumps in the observed field around M15 could be due to 
unidentified background galaxy clusters, 
although significant contamination by the clustered background objects
seems a priori unlikely in the field~\citep[cf.][]{Ode01}.

M30 is located on 7.1 kpc from the Galactic center. The cluster is 
one of the most metal-poor globular clusters in the Milky Way with [Fe/H] $=-2.12$,
and it has a very high concentration of $c=2.5$ ~\citep{Har96}.
Extended extratidal features on the configuration of stars surrounding M30
have not yet been reported.
Analyses of spatial surface density maps of stars in the vicinity of M30
revealed flocculent overdensity features with tails around the cluster 
that extend to about $2r_t$ and beyond. 
The radial surface density profiles of stars have a break at $\sim0.6r_t$ for M30,
and the surface density excess seems to extend to $\sim2r_t$. 
The apparent east-west extension of the extratidal tails 
is likely associated with the direction of the Galactic center and anticenter.
This is consistent with the extratidal overdensity property
appearing in the radial surface density profiles
and the azimuthal number density profiles for stars surrounding M30.
Although tails associated with the direction of proper motion do not clearly show
in the spatial surface density maps, the radial density profiles
in $0.6r_t\lesssim r \lesssim 2r_t$ for M30 shows relatively shallower slopes
with a power law in the direction of the cluster's spatial orbit.
We note, however, the visibility of such tails around M30
depends on the orientation of the cluster's orbit
to our line of sight, i.e., tails could not be observed if we were 
situated in the plane of a cluster's orbit.
The background features of galaxies and dusts are not likely to disturb
the observed flocculent overdensity features of tails around M30.
Indeed, the observed field is polluted only by two Abell clusters.
The dust extinction is low toward this cluster (E$(B-V)=0.03$) and 
the fluctuation of the dust emission is low as traced by the IRAS 100$\mu$m map.

\begin{figure}
\centering
\includegraphics[width=0.7\textwidth]{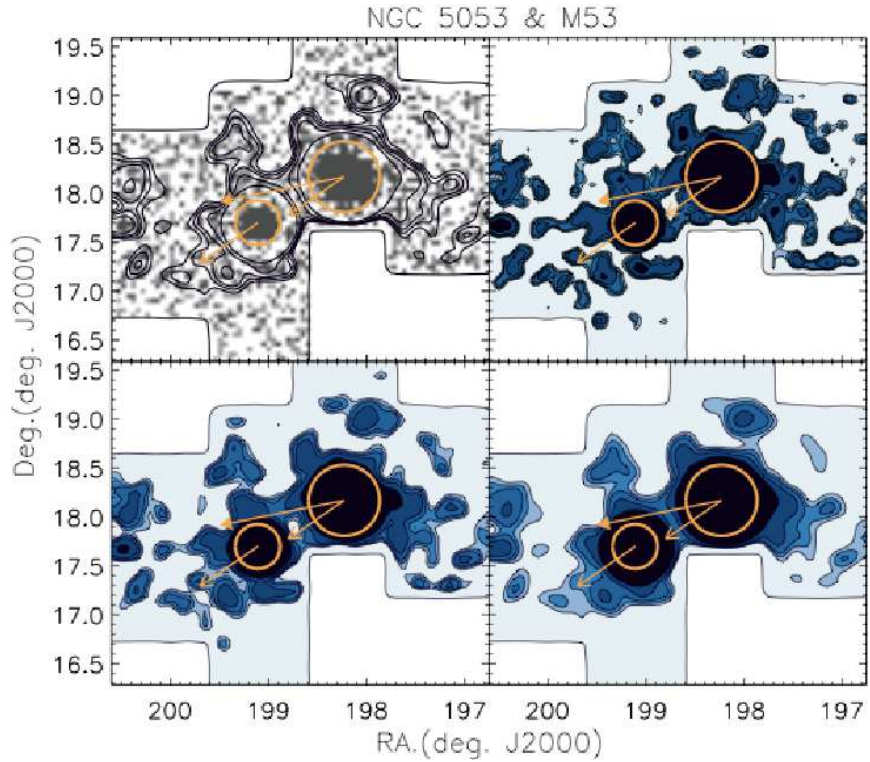}
\caption{The raw star count map around M53 and NGC 5053 and the surface density
contour maps smoothed by Gaussian kernel values of $0.125^{\circ}, 0.175^{\circ}$, 
and $0.240^{\circ}$. The contour levels are the backgound level, $2\sigma,4\sigma,
8\sigma$, and $20\sigma$ above the background level.
The circles indicate tidal radii of M53 and NGC 5053.
The long arrow from M53 indicates the direction of the cluster's proper motion.
The short arrows represent the direction of the Galactic center.
\label{N5053M53contour}}
\end{figure}

\begin{figure}
\centering
\includegraphics[width=0.3\textwidth]{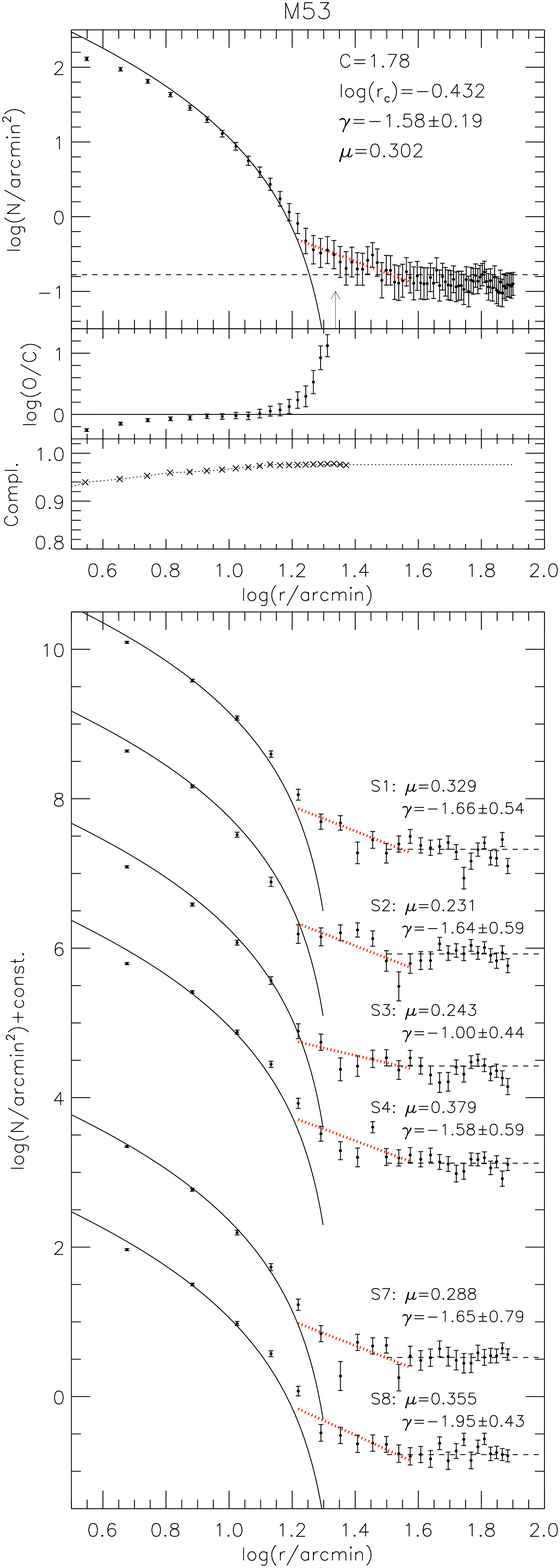}
\caption{{\it Upper} : Radial surface density profile of M53 with a theoretical King 
model of $c=1.78$ (solid curve), residuals after subtracting the values predicted by the King model,
and radial completeness and its trend (dotted line).
The profile within the detected overdensity region, $\sim 0.7r_t\leq r\leq \sim 1.6r_t$, 
is represented by a power-law, i.e., a dotted straight line in logarithmic scale
with a slope of $\gamma=-1.58\pm0.19$.
The mean number density of stars in the overdensity region is estimated
as $\mu=0.302$ per arcminute square. 
The dashed horizontal line indicates the 
background density level, and an arrow represents the tidal radius of M53.
{\it Lower} : Radial surface density profiles of six different angular sections.
For clarity, the surface densities from S1, S2, S3, S4, S7 and S8 
are given zero-point offsets with values of 
const.= 8.1, 6.7, 5.2, 3.9, 1.3, 0.0.
Dotted lines indicate the power-law fit to the overdensity region with $\gamma$ as a slope.
$\mu$ is the mean density in the overdensity region.\label{M53king}}
\end{figure}

\begin{figure}
\centering
\includegraphics[width=0.7\textwidth]{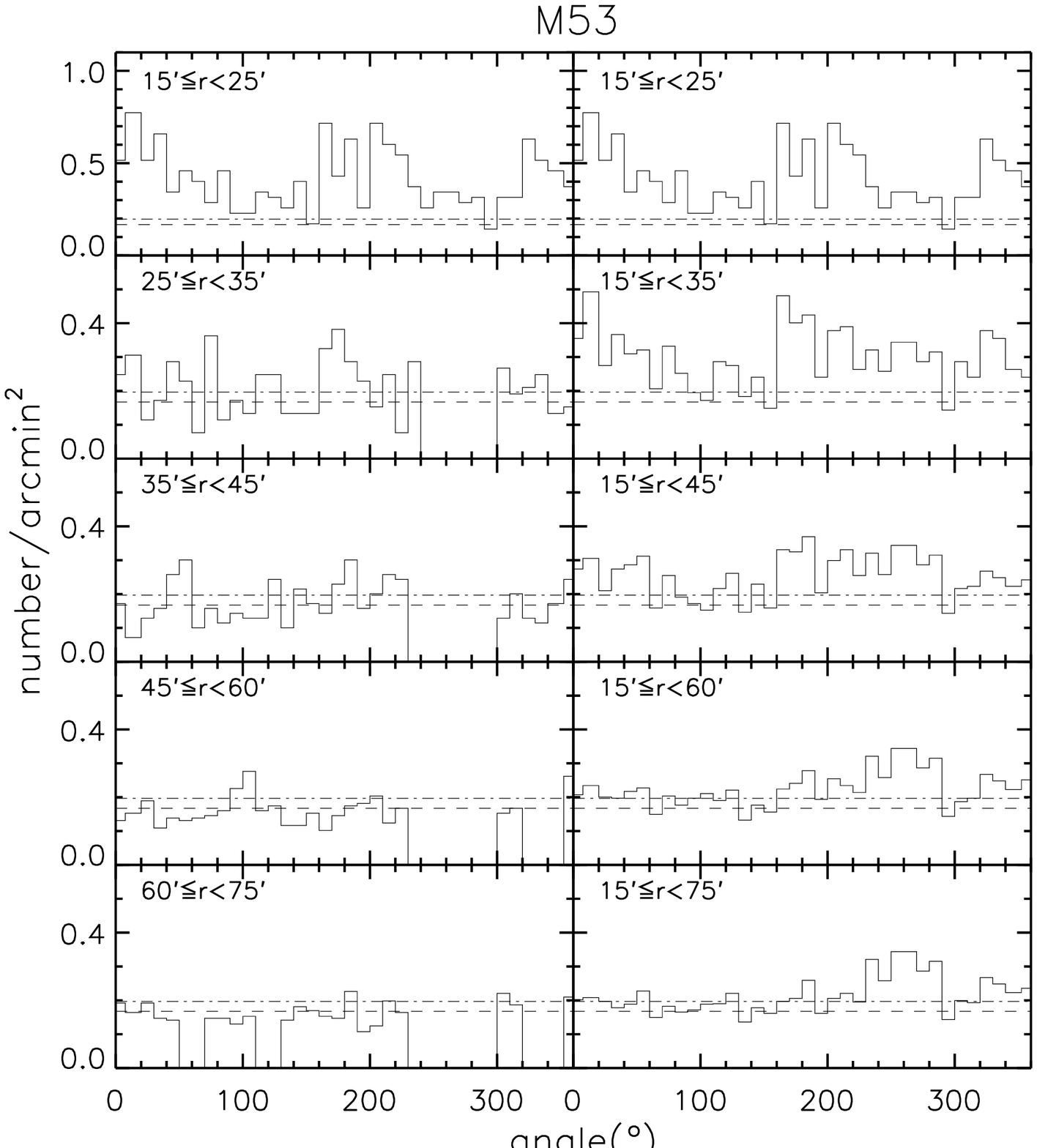}
\caption{The annular (left panels) and radially cumulative (right panels)
azimuthal number density profiles of M53 with an angular bin size of $10^{\circ}$.
The radial bins are set to be $10^{'}$ for the inner three annuli and $15^{'}$ for the 
outer two annuli.
The blank bins correspond to the areas without photometric data and 
the area within the tidal radius of the neighbor cluster NGC 5053.
The horizontal dashed lines are the determined backgound density level and 
dashed dotted lines are the level of 1$\sigma$ deviation above background level.
\label{M53azim}}
\end{figure}

\begin{figure}
\centering
\includegraphics[width=0.3\textwidth]{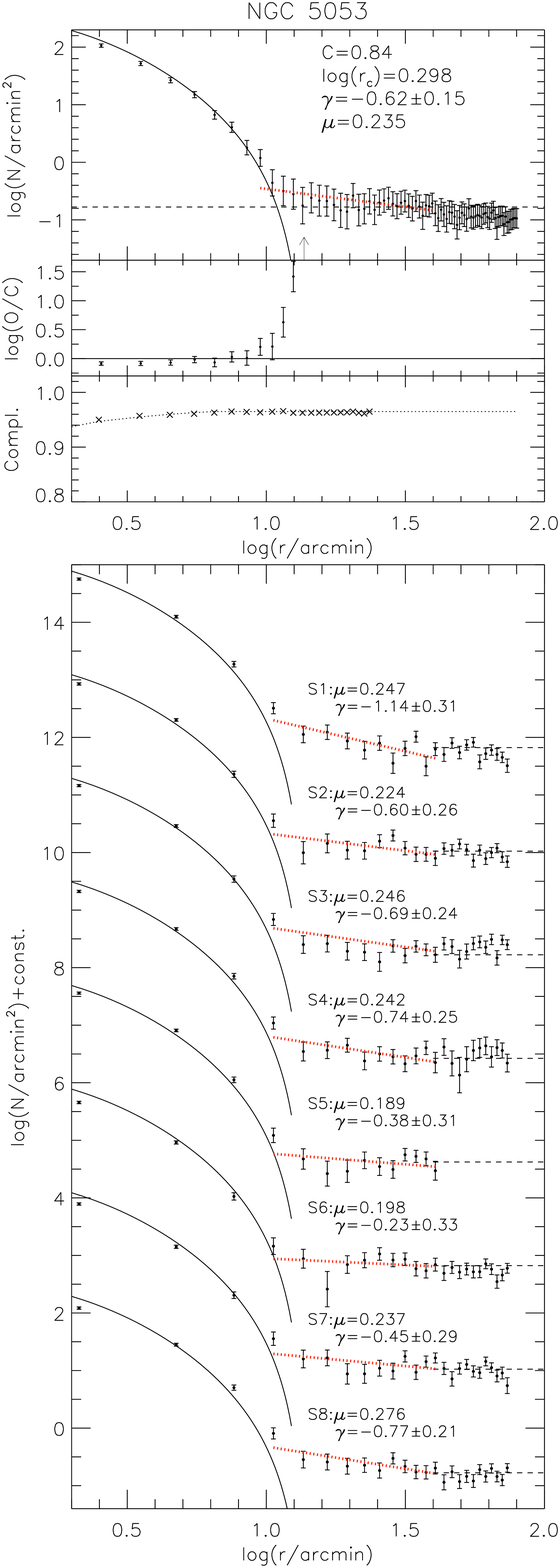}
\caption{{\it Upper} : Radial surface density profile of NGC 5053 with a theoretical King 
model of $c=0.84$ (solid curve), residuals after subtracting the values 
predicted by the King model, and radial completeness and its trend (dotted line).
The profile within the detected overdensity region, $\sim 0.7r_t\leq r\leq \sim 2.5r_t$, 
is represented by a power-law, i.e., a dotted straight line in logarithmic scale
with a slope of $\gamma=-0.62\pm0.15$.
The mean number density of stars in the overdensity region is estimated
as $\mu=0.235$ per arcminute square. 
The dashed horizontal line indicates the 
background density level, and an arrow represents the tidal radius of NGC 5053.
{\it Lower} : Radial surface density profiles of eight different angular sections (S1$\sim$S8).
For clarity, the surface densities from S1 to S8 
are given zero-point offsets with values of 
const.= 12.6, 10.8, 9.0, 7.2, 5.4, 3.6, 1.8, 0.0.
Dotted lines indicate the power-law fit to the overdensity region with $\gamma$ as a slope.
$\mu$ is the mean density in the overdensity region.
\label{N5053king}}
\end{figure}

\begin{figure}
\centering
\includegraphics[width=0.7\textwidth]{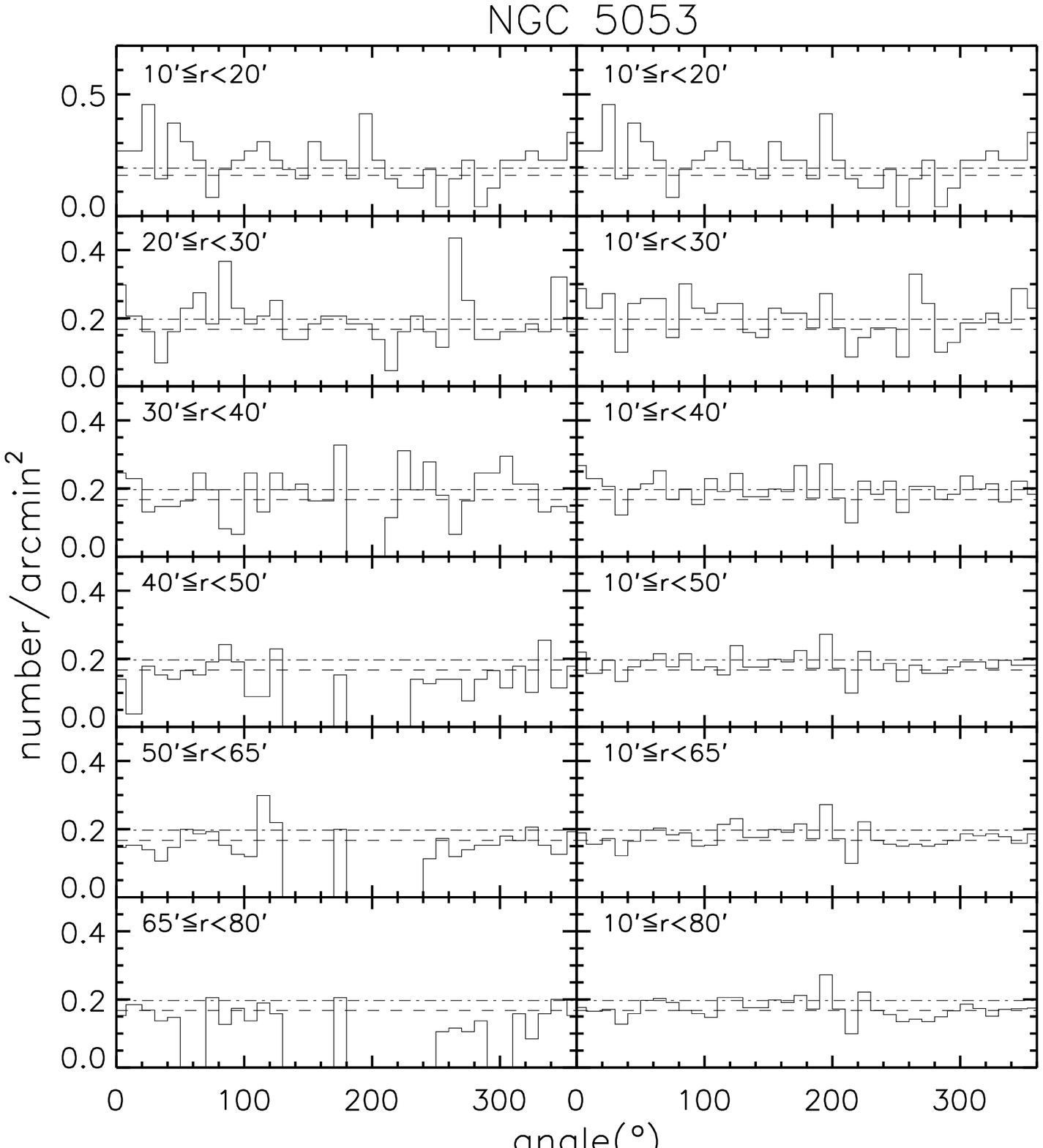}
\caption{The annular (left panels) and radially cumulative (right panels)
azimuthal number density profiles of NGC 5053 with an angular bin size of $10^{\circ}$.
The radial bins are set to be $10^{'}$ for the inner four annuli and $15^{'}$ for the 
outer two annuli.
The blank bins correspond to the areas without photometric data and 
the area within the tidal radius of the neighbor cluster M53.
The horizontal dashed lines are the determined backgound density level and 
dashed dotted lines are the level of 1$\sigma$ deviation above background level.
\label{N5053azim}}
\end{figure}

The clusters M53 and NGC 5053 are close to each other,
with a projected angular distance of about $1^{\circ}$, 
so they constitute an unusual astronomical pair in the outer halo of the Milky Way. 
These two clusters are the most metal-poor halo globular clusters
with similar metallicities, i.e., [Fe/H] $= -1.99$ for M53 and [Fe/H] $=-2.29$ for NGC 5053.
Also they have the same relative age as M92~\citep{Hea91,Rey98} 
and a nearly the same distance modulus.
The diffuse cluster NGC 5053 has a tidal radius of about $13.67^{'}$ with a very low 
concentration parameter $c=0.84$, while M53 has a rather higher concentration $c=1.78$ 
and a tidal radius of $21.75^{'}$~\citep{Har96}.
In this study, we identified extratidal features on the spatial configuration of 
selected stars in the vicinity of two clusters M53 and NGC 5053.
The observed tails  likely trace the clusters' orbits and the direction to the 
Galactic center.
As new findings, we detected features of clumps and ripples surrounding the clusters,
and a tidal bridge-like feature and  
an envelope structure around the multiple globular cluster system.
Here, we also note that~\citet{Lau06} reported a discovery of a $6^{\circ}$ tidal stream 
extending from NGC 5053 in the southwest direction, using the SDSS data from which they 
removed stars within the tidal radius of M53 from the analysis.
Unfortunately, it is difficult to confirm the extended stream in the surface density maps of 
Figure~\ref{N5053M53contour}, because our observations do not contain photometric data in
the southwest direction of the M53 and NGC 5053 system.
Such spatial stellar configuration features
might be a result of tidal interaction or merging between the two clusters.
Indeed, the structures of clumps and ripples are predicted from recent numerical simulations 
for the interaction of two globular clusters~\citep[e.g.,][]{Mio06}.
Furthermore, features such as a tidal bridge and an envelope 
in a multiple globular cluster system are also predicted from the numerical simulations 
of the possible merging of a set of globular clusters~\citep[e.g.,][]{Cap08}.
From an observational point of view, ~\citet{Van96} suggested that 
a merging of clusters would occur more preferentially in dwarf galaxies 
with low velocity dispersions of the cluster system in the range of 
6.5$\sim$11 $kms^{-1}$~\citep{Mat96} rather than in the Milky Way.

To test the contamination of background galaxies in the field of M53 and NGC 5053, 
we analyzed the spatial distribution of non-pointlike SDSS sources.
For the extracted SDSS sources, we transformed the SDSS $g^{'}r^{'}i^{'}$ magnitude system 
into $c_1$ and $c_2$ indices, using the Equation (1). Sources falling
into the masks in $(c_2,i^{'})$ and $(c_1,i^{'})$  planes, which were used to select
cluster member stars, were applied to the sample of background galaxies.
In Figure~\ref{N5053M53Gal}, we show a smoothed surface density map of 
the selected background galaxies, overlaid with the surface density contours 
of cluster stars of Figure~\ref{N5053M53contour}.
Filled triangles indicate two Abell clusters in the observed field.
As can be seen in Figure~\ref{N5053M53Gal}, we found no strong correlation between 
the pattern of the surface density variation of the background galaxies
and the observed spatial configuration of stars around M53 and NGC 5053.
Also note that the dust extinction is low toward the clusters (i.e., E$(B-V)=0.02$ 
for M53 and E$(B-V)=0.04$ for NGC 5053), and the variation in dust emission 
through the observed field is low as traced by the IRAS 100 $\mu$m map.

According to photometric results in this paper, it could be suggestive that
the clusters M53 and NGC 5053 may be interacting or have interacted in the past.
In the kinematic point of view, however,
it would not be airtight that they originated with a same common dwarf spheroidal parent.
The radial velocity of M53 is $-79.1 \pm 4.1$ $km$ $s^{-1}$,
but it is $+44.0 \pm 0.4$ $km$ $s^{-1}$ for NGC 5053 ~\citep{Har96}.
In this case, the clusters might have a common space orbit to the similar direction,
only if the tangential velocities are in same directions with
exorbitantly larger values than their absolute values of radial velocities.
The tangential velocity of M53 is estimated to be $\mu=43.1\pm84.0$ $km$ $s^{-1}$
by using values of the proper motion, i.e., $\mu_{\alpha}\cos\delta=0.50\pm1.00$ mas yr$^{-1}$ and
$\mu_{\delta}=-0.10\pm1.00$ mas yr$^{-1}$  ~\citep{Din99}.
Even if we are taking into account of the large measurement error,
the calculated tangential velocity of M53 is 
merely comparable to the absolute value of the radial velocity. 
Unfortunately, we can not estimate the tangential velocity of NGC 5053 
because the proper motion has not yet been reported.
Thus, the current kinematic information about radial velocities and inaccurate tangential motions
of M53 and NGC 5053 
makes it still hard to arrange if the clusters came from 
a common dwarf and were on similar paths through the Galactic halo. 
Also, their distance moduli, $(m-M)_v=16.31$ for M53 and 
$16.19$ for NGC 5053 ~\citep{Har96}, are still separated by $\sim1.4$ kpc, 
a significant distance in the halo of the Milky Way
compared with their tidal radii 112.7 pc for M53 and 66.8 pc for NGC 5053 ~\citep{Har96}. 
Nevertheless, 
it should be noted that radial velocities of M53 and NGC 5053
are commensurate with those of the other clusters aligned 
in the Magellanic plane ~\citep[see Figure 4c in][]{Yoo02}.
Indeed, radial velocities of the globular clusters in the plane show a 
sinusoidal pattern of Keplerian motion as they are plotted
against the orbital longitude measured along the Magellanic plane.
This is possibly indicative of a common origin of the clusters.

\begin{figure}
\centering
\includegraphics[width=0.7\textwidth]{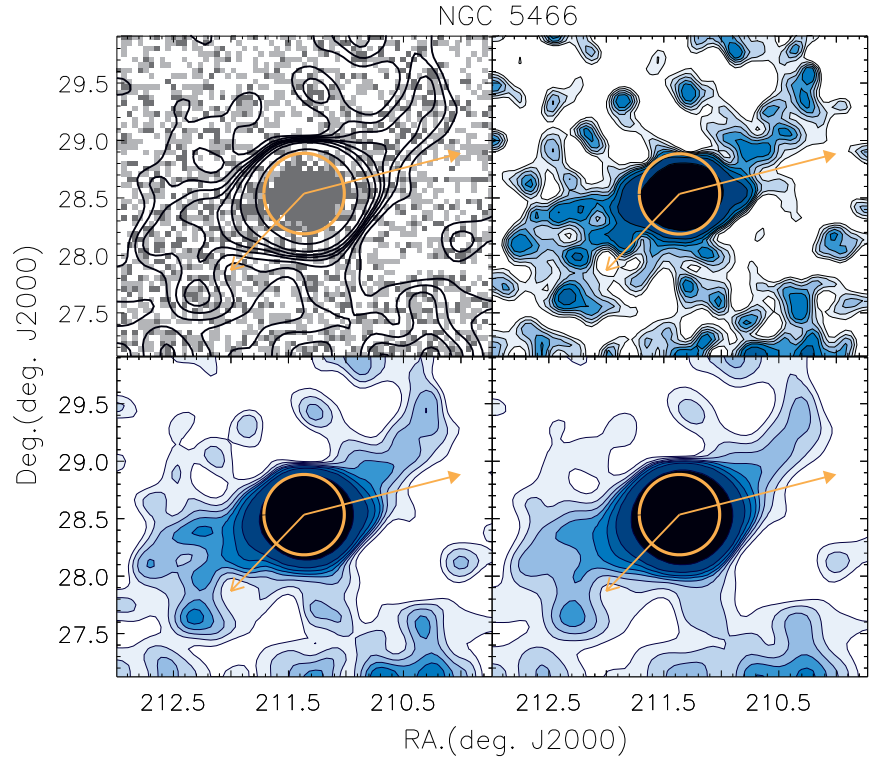}
\caption{The raw star count map around NGC 5466 and the surface density
contour maps smoothed by Gaussian kernel values of $0.168^{\circ}, 0.264^{\circ}$, 
and $0.312^{\circ}$.
The contour levels are the backgound level, $1\sigma, 2\sigma,
3\sigma, 4\sigma, 6\sigma, 10\sigma$, and $40\sigma$ above the background level.
The circle indicates a tidal radius of NGC 5466 ~\citep{Leh97}.
The short and long arrows represent the direction of the Galactic center
and the cluster's proper motion, respectively.\label{N5466con}}
\end{figure}

\begin{figure}
\centering
\includegraphics[width=0.3\textwidth]{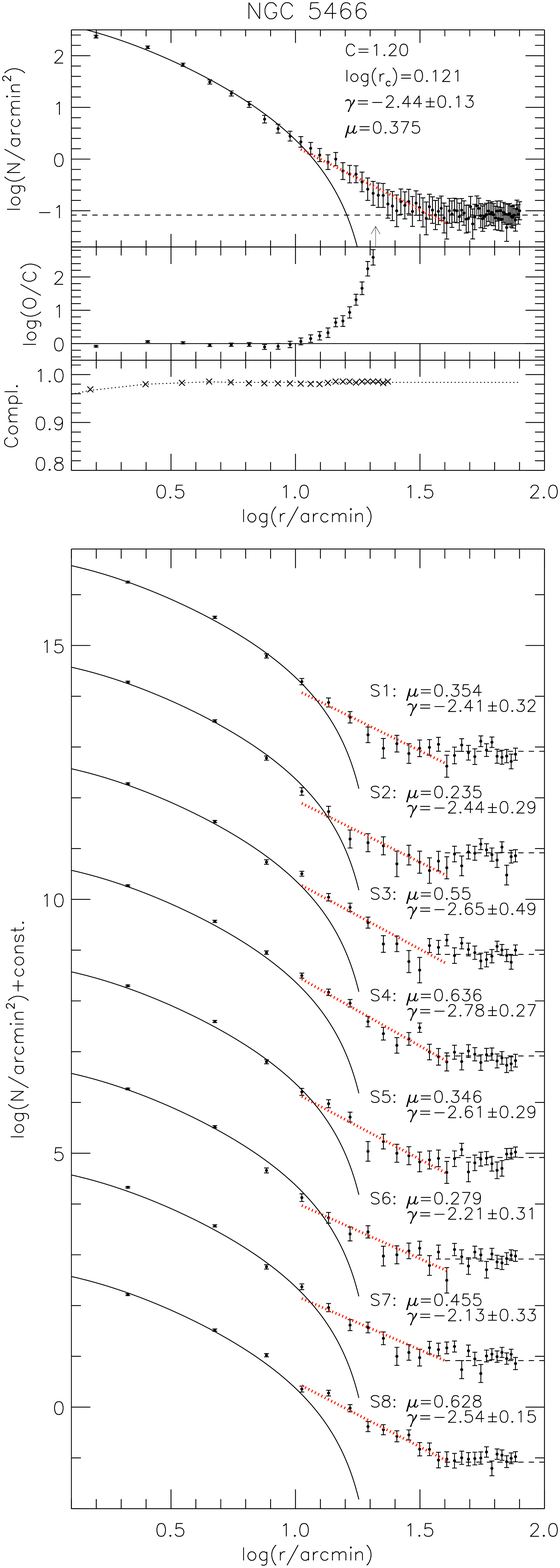}
\caption{{\it Upper} : Radial surface density profile of NGC 5466 with a theoretical King 
model of $c=1.20$ (solid curve), residuals after subtracting the values 
predicted by the King model, and radial completeness and its trend (dotted line).
The profile within the detected overdensity region, $\sim 0.5r_t\leq r\leq \sim 2r_t$, 
is represented by a power-law, i.e., a dotted straight line in logarithmic scale
with a slope of $\gamma=-2.44\pm0.13$.
The mean number density of stars in the overdensity region is estimated
as $\mu=0.375$ per arcminute square. 
The dashed horizontal line indicates the 
background density level, and an arrow represents the tidal radius of NGC 5466.
{\it Lower} : Radial surface density profiles of eight different angular sections (S1$\sim$S8).
For clarity, the surface densities from S1 to S8 
are given zero-point offsets with values of 
const.= 14.0, 12.0, 10.0, 8.0, 6.0, 4.0, 2.0, 0.0.
Dotted lines indicate the power-law fit to the overdensity region with $\gamma$ as a slope.
$\mu$ is the mean density in the overdensity region.
\label{N5466king}}
\end{figure}

\begin{figure}
\centering
\includegraphics[width=0.7\textwidth]{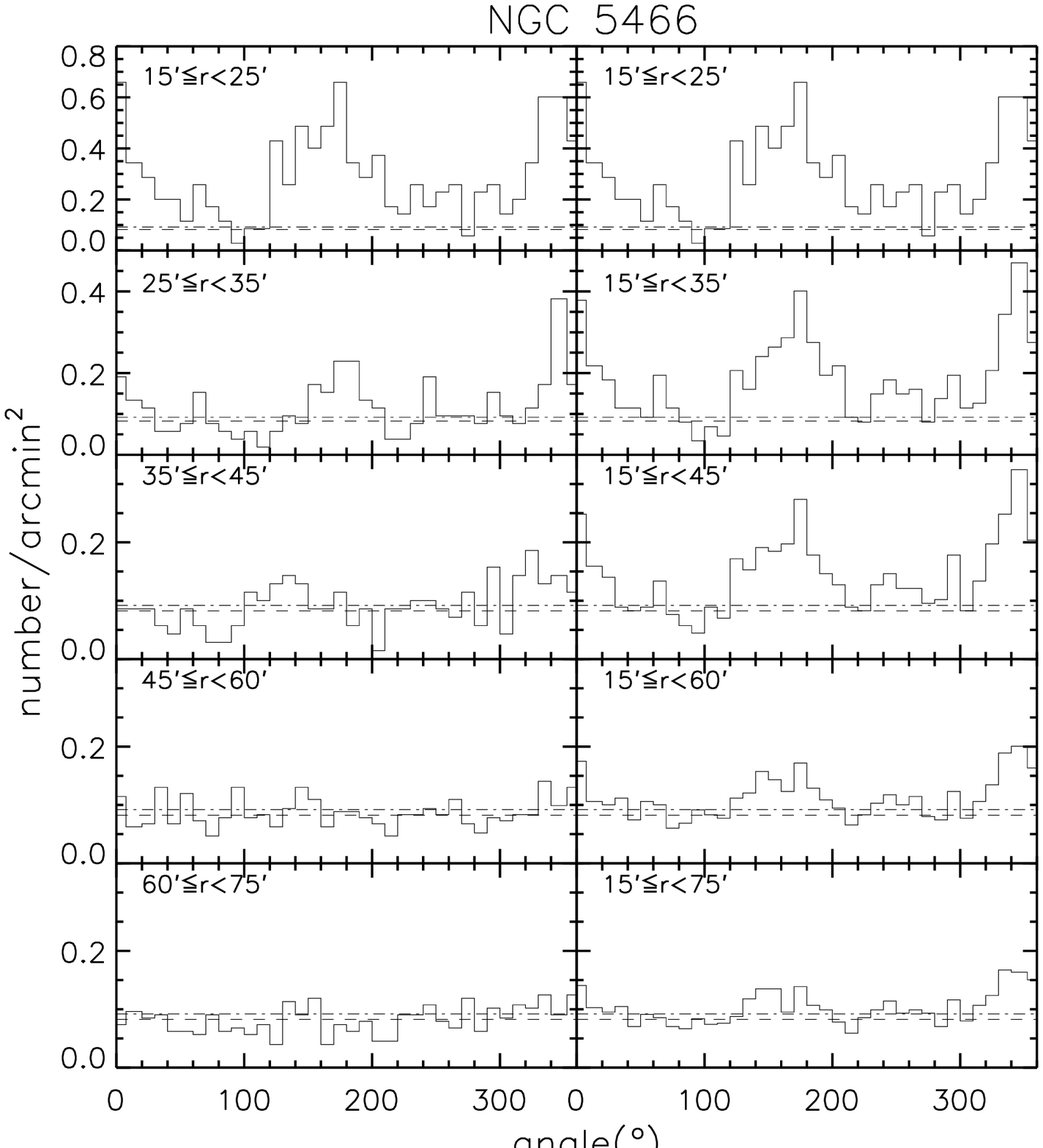}
\caption{The annular (left panels) and radially cumulative (right panels)
azimuthal number density profiles of NGC 5466 with an angular bin size of $10^{\circ}$.
The radial bins are set to be $10^{'}$ for the inner three annuli and $15^{'}$ for the 
outer two annuli.
The horizontal dashed lines are the determined backgound density level and 
dashed dotted lines are the 1$\sigma$ deviation above background level.
\label{N5466azim}}
\end{figure}

\begin{figure}
\centering
\includegraphics[width=0.7\textwidth]{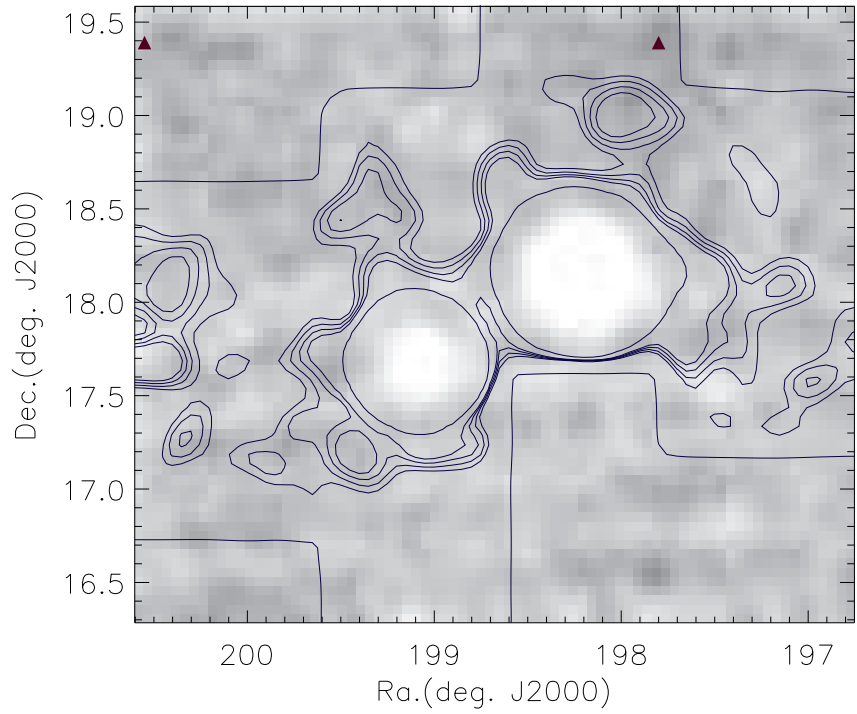}
\caption{The gray scale map for the surface number density of background galaxies 
around M53 and NGC 5053.
Background galaxies were selected by using non-pointlike SDSS sources. 
Filled triangles indicate Abell clusters in the observed field.
The overlaid contour levels show the spatial configuration of selected stars
surrounding two clusters, M53 and NGC 5053.
\label{N5053M53Gal}}
\end{figure}

The cluster NGC 5466 is at a 16.2 kpc distance from the Galactic center. It is one of the
low-metallicity clusters with [Fe/H] $=-2.22$, and it
has a low concentration with $c=1.2$ ~\citep{Har96}.
~\citet{Leh97} found evidence of tidal tails around NGC 5466 
in a radial density profile of the cluster.
Recently, ~\citet{Bel06b} observed tidal tails of NGC 5466, stretching about $2^{\circ}$ 
in apposite directions. Using SDSS photometric data, 
~\citet{Gril06} found evidence for a much larger extension of the tidal tails, 
with the leading arm over $\sim30^{\circ}$ and the trailing arm extending at 
least $15^{\circ}$.
From numerical simulations, ~\citet{Fel07} reproduced the tidal tails around 
cluster NGC 5466. In our study, analyses for spatial configuration of stars around NGC 5466 
have indicated an overdensity feature extending outside the tidal radius of the cluster.
In the spatial surface density maps in Figure~\ref{N5466con}, we have
found an S-shape extratidal feature of tails stretching out symmetrically 
over $\sim2^{\circ}$ in the observed field. 
The observed radial density profiles of NGC 5466, as shown in Figure~\ref{N5466king}, 
show a departure from the King model, with a break at $\sim0.5r_t$
extending the overdensity feature out to $\sim2r_t$ of the cluster. 
Azimuthal number density profiles in Figure~\ref{N5466azim} confirmed 
the orientation of the two tails around NGC 5466, which is likely
consistent with the cluster's orbit inferred from the cluster's proper motion.
The results are reconfirmation of the tidal tails of NGC 5466 stretching 
as much as $45^{\circ}$ on the sky as reported by ~\citet{Bel06b} and ~\citet{Gril06}.

\begin{figure}
\centering
\includegraphics[width=0.7\textwidth]{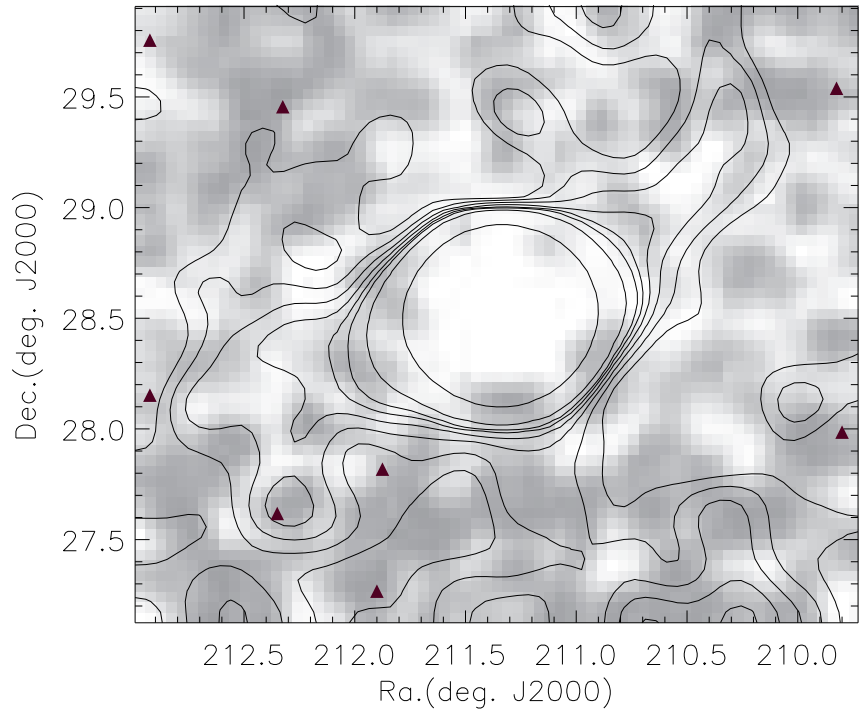}
\caption{The gray scale map for the surface number density of background galaxies around NGC 5466.
Background galaxies were selected by using non-pointlike SDSS sources. 
Filled triangles indicate eight Abell clusters in the observed field.
The overlaid contour levels show the spatial configuration of selected stars
surrounding NGC 5466.
\label{N5466Gal}}
\end{figure}

Similar to the case of M53 and NGC 5053, 
we checked contamination of background galaxies in the field of NGC 5466
by analyzing the spatial distribution of non-pointlike SDSS sources.
Figure~\ref{N5466Gal} shows a smoothed surface density map of 
the selected background galaxies overlaid with the surface density contours 
of cluster stars from Figure~\ref{N5466con}.
Filled triangles indicate eight Abell clusters in the observed field.
As shown in Figure~\ref{N5466Gal}, the pattern of density variations in the selected galaxy 
sample does not seem to be correlated with the location of the tidal tails around NGC 5466.
However, the spatial configuration of the selected background galaxies
in the southern area of the observed field is likely correlated with 
the clumpy overdensity feature of the surface density maps of 
selected cluster member stars of Figure~\ref{N5466con}.
The observed field is not polluted  by 
strong dust extinction (E$(B-V)=0.00$).

The other two low-metallicity clusters M68 and M92 in the Magellanic plane, 
for which we did not secure the wide-field deep photometric data from the present observations,
are located on 10.1 kpc and 9.6 kpc from the Galactic center. 
Indications of extratidal extensions of stars around the clusters incidently
have been overlooked in a few previous studies. 
~\citet{Tes00} used the DPOSS plates to show an circular structure 
of extratidal halo around M92 extending out to $\sim30^{'}$ from the cluster center. 
Also, ~\citet{Lee03} used CFHT mosaic CCD imaging to discover a presence of 
extratidal materials in the radial density profile of M92 and 
a weak tidal halo around the cluster. 
In addition, ~\citet{Gri95} found an indication 
for extratidal overdensity feature around M68 using photographic plates taken at the UKST. 
Nevertheless, there have been little evidences for extratidal tails 
toward any preferred directions on the spatial stellar distributions around both of the clusters 
because of the lower spatial resolutions and 
the brighter limiting magnitudes of the photographic and CCD photometry. 
Thus we note the need for deeper wide-field CCD photometry to investigate
spatial structure of the extended extratidal tails around both of the clusters.

The spatial configuration of stars tidally stripped from a parent stellar system
of a present globular cluster contains direct information of the dynamical evolution 
that the cluster has experienced.
Along with internal dynamical processes,
external perturbations, such as disk and bulge shocks by the Galaxy,
could accelerate the cluster's dynamical evolution
~\citep[e.g.,][]{Spi71,Gne97}.
With respect to external dynamical evolution,
three clusters, M15, NGC 5053 and NGC 5466 among our five targets might have 
experienced strong interactions with the Galaxy in the form of disk and bulge shocks,
as indicated by the ratios of destruction rates with 
$v_{tot}/v_{evap}=3.3, 130$, and $160$, respectively, for each cluster~\citep{Gne97}.
Here, $v_{tot}$ and $v_{evap}$ are the total destruction rate and the evaporation rate
per Hubble time, respectively.
Furthermore, ~\citet{Leo00} suggested that tidal tails usually stretch
into the Galactic center, as the effect of both bulge and disk shock or the bulge 
shock to the cluster is much stronger than disk shock for the cluster.
This implies that tidal tails stretching into the Galactic center and anticenter
around the three clusters M15, NGC 5053 and NGC 5466 might be a result of 
both the bulge and disk shock of the Galaxy.
Two other clusters, M30 and M53, are not expected to suffer strong
gravitational shocks by the Galaxy, with $v_{tot}/v_{evap}=1.0$.
This would be a reason why the tidal tail feature of M30 is not well defined
in the direction of the Galactic center.
Also, the tidal bridge-like feature and an envelope structure around M53
may not be the results of gravitational shocks of the Galaxy
but the interaction with the companion cluster NGC 5053. 
In addition, apparent extratidal extensions around the other two globular clusters 
M68 and M92 with $v_{tot}/v_{evap}=1.37$ and $1.00$, respectively,
may not be the results of suffering strong gravitational shocks by the Galaxy.
In spite of such a dependency on the Galactic potential
to the dynamical structure of the observed tidal tails,
we found here that the feature of extratidal extensions around the target clusters
is unlikely to have any distinctive correlation with 
the Galactocentric distance of each cluster.

\begin{figure}
\centering
\includegraphics[width=0.6\textwidth]{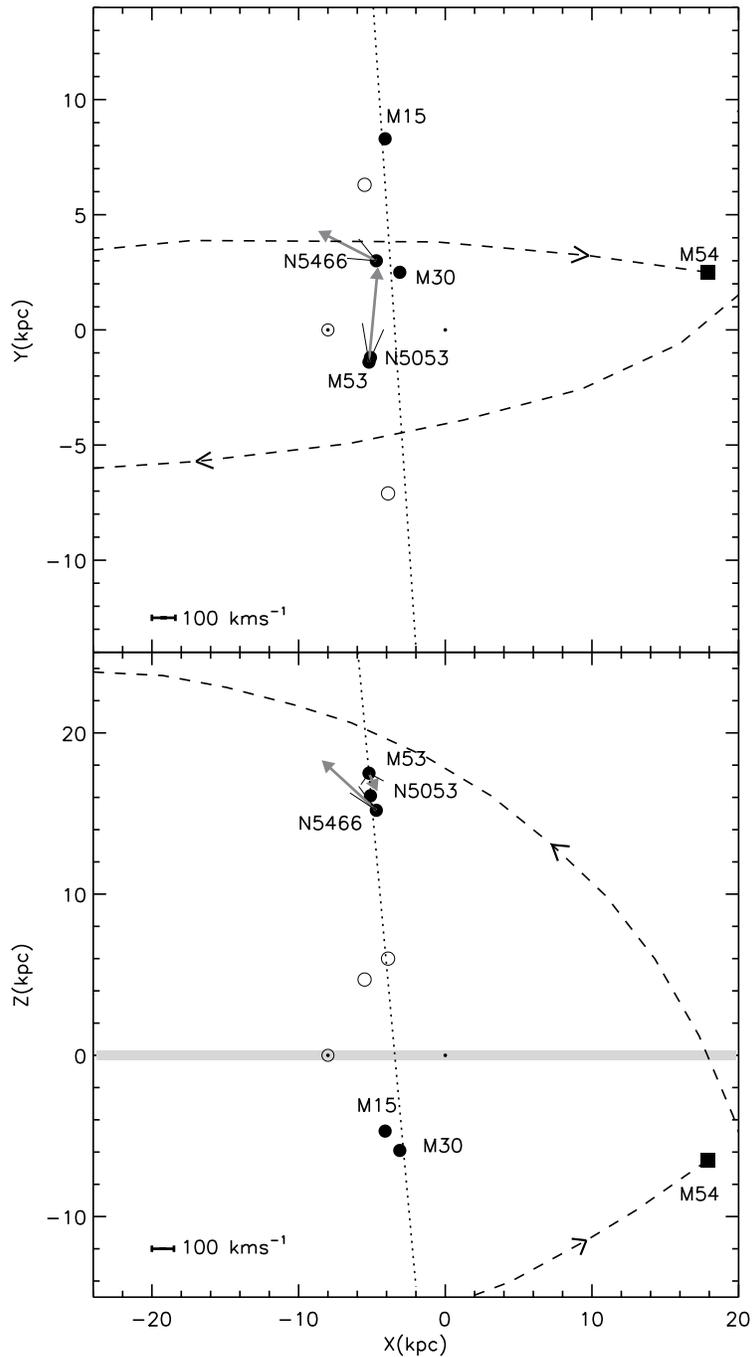}
\caption{The spatial distribution of seven low-metallicity globular clusters in the Galactocentric coordinate. 
The Galactic plane is drawn in $X-Z$ plane, and the 
Sun (indicated by the circle with a dot) is at (-8,0,0) kpc. 
The filled circles are five target clusters in this study, 
while two open circles represent M68 and M92.
Space motion vectors of M53 and NGC 5466 are indicated by thick arrows
with the uncertainties.
The filled square marks the Sagittarius globular cluster M54.
The dotted line is the Magellanic plane ~\citep{Yoo02}, and
the dashed line is the Sgr orbit ~\citep{Iba98} since the present time
to $\sim1$ Gyr ago.
\label{psmv}}
\end{figure}

On the other hands, many numerical simulations of tidal tails performed
over the past decade ~\citep[i.e.,][]{Com99,Yim02,Cap05,Mon07} have suggested
that the stars dissolved from a globular cluster by the Galactic tidal force 
slowly drift along the cluster's orbit due to the orbital motion and the coriolis acceleration, 
and then eventually form a dense tidal tail parallel to the cluster's orbit.
In this paper, we found that all of the observed low-metallicity clusters have shown 
possible extratidal extensions to the direction of the clusters' orbits, 
which are inferred from their proper motions.
Indeed, the five target globular clusters are assumed to lie on one
spatial orbital plane in the outer Galactic halo, and their kinematic motions place also  
almost on this plane~\citep{Yoo02}, which are considered as a result of accretion
of dwarf galaxies. Based on a comparison of population and kinematics, 
~\citet{Yoo02} suggested that LMC
is the possible parent galaxy of these halo globular clusters.
However, recent studies ~\citep{Pal02,Bel03,Mart04} have stated that 
M53, NGC 5053 and NGC 5466 are possible former members of the Sgr galaxy.
To examine a possible association of the clusters with
the Sgr orbit or Magellanic debris plane,
we mark the three clusters on the planes formed by the rectangular Galactocentric coordinates 
with $X, Y,$ and $Z$ in kiloparsecs, as shown in Figure ~\ref{psmv}.
The Magellanic debris plane ~\citep{Yoo02} and the Sgr orbit ~\citep{Iba98} 
are also overlaid in Figure ~\ref{psmv}.
We also plotted the space motion vectors of two clusters M53 and NGC 5466 
with the uncertainties in directions, 
in order to see their velocity coherence to the two orbital planes.
Space motion vector of NGC 5053 is not plotted, since
the proper motion of the cluster is not available.
As shown in Figure ~\ref{psmv}, M53 is likely to move in a direction along the
Magellanic plane.  
The space motion of M53 is perpendicular to the Sgr orbit in $X-Y$ plane, 
and it is directionally opposite in $X-Z$ plane to the orbital direction of the Sgr dwarf.
The space motion vector of NGC 5466 in $X-Z$ plane seems to be parallel 
to the direction of the Sgr orbit,
but it is not conclusive that NGC 5466 is moving along the Sgr orbit.
The space motion of NGC 5466 in $X-Y$ plane is against the Sgr orbit,
and it seems to move in a direction along to the Magellanic plane within
the uncertainty order of $\sim2.5\sigma$ ~\citep[e.g.,][]{Yoo02}.
In addition, NGC 5466 is closer in space to the Magellanic plane than to the Sgr orbit.
NGC 5053 lies remarkably closer to the Magellanic plane than to the Sgr orbit,
while it has no information about the space motion.
These lead us to conclude that the three clusters M53, NGC 5053, and NGC 5466, are 
potentially associated with the Magellanic plane rather than the Sgr orbit.

Under these observational conditions, 
tidal tails for the five target globular clusters are expected to have
a common feature of kinematic orbital properties, which would provide information 
about their parent dwarf galaxies. Although the spatial extensions and morphological shapes
exhibited different regimes of developments for the observed extratidal tails,
all the observed target clusters presented evidences of tidal tails tracing
their orbits, as inferred from the projected directions of their proper motions.
Thus, it appears that the tidal tails to the cluster's orbit
for five target metal poor halo clusters are indeed evidences for an accreted origin from 
satellite system.
Also, these observational results of extratidal tail features around the five target clusters
provide further evidences of the merging scenario of the formation for the Galactic halo,
and in addition an important constraint on the potential in the Galactic halo ~\citep{Ode09}.

Data for more accurate proper motions of target clusters and
radial velocities of stars in the observed extratidal structures
are essential to verify their association with a stream in the Galactic halo.
Furthermore, larger area coverage with accurate deep photometric data 
will also enable us to better constrain the orbits of the target clusters
and cluster's member stars in the extratidal structures.
With information about the spatial positions, 
radial velocities and proper motions of stars in the tidal tails, 
we could integrate their orbits backwards in the Galactic potential 
and eventually ascertain the origin of the globular clusters in the Galactic halo.

\acknowledgments
We are grateful to an anonymous referee for a detailed report that greatly
improved this paper.
This work has been supported by the Korea Research Foundation Grant funded by
the Korea Government (KRF 2007-313-C00321), for which we are grateful.

%
%
%
%
%
%
%
%
%
\clearpage
%

%
%
%
%
%
%
%
%
%

\clearpage
\begin{deluxetable}{cccccccccc}
\tabletypesize{\scriptsize}
\tablewidth{0pt}
\tablecaption{Basic data for five target globular clusters\label{posi}}
\tablehead{
\colhead{Target} & \colhead{$\alpha$} & \colhead{$\delta$} &
\colhead{l} & \colhead{b} & \colhead{$R_{sun}$} & \colhead{$R_{GC}$} & \colhead{$r_c$} & 
\colhead{$r_t$} & \colhead{[Fe/H]}\\
 & \colhead{(J2000)} & \colhead{(J2000)} & \colhead{(deg)} & \colhead{(deg)} &
 \colhead{(kpc)} & \colhead{(kpc)} & \colhead{($'$)} & \colhead{($'$)}
}
\startdata
M15 & 21:29:58.3 & 12:10:01 & 65.01 & --27.31 & 10.3 & 10.4 & 0.07 & 21.5 & --2.26\\
M30 & 21:40:22.0 & --23:10:45 & 27.18 & --46.83 & 8.0 & 7.1 & 0.06 & 18.34 & --2.12\\
M53 & 13:12:55.3 & 18:10:09 & 332.96 & 79.76 & 17.8 & 18.3 & 0.36 & 21.75 & --1.99\\
NGC 5053 & 13:16:27.0 & 17:41:53 & 335.69 & 78.94 & 16.4 & 16.9 & 1.98 & 13.67 & --2.29\\
NGC 5466 & 14:05:27.3 & 28:32:04 & 42.15 & 73.59 & 15.9 & 16.2 & 1.20 & 20.98 & --2.22\\
\enddata
\tablecomments{ 
$R_{sun}$ and $R_{GC}$ represent distances from the Sun and the Galactic center, 
respectively. $r_c$ and $r_t$ are the core radius and tidal radius of a cluster.
Data are from ~\citet{Har96} except for core radius and tidal radius of NGC 5466, 
which are from ~\citet{Leh97}.
}
\end{deluxetable}
%
\begin{deluxetable}{ccccc}
\tabletypesize{\scriptsize}
\tablewidth{0pt}
\tablecaption{Journal of Observations\label{log}}
\tablehead{
\colhead{Target} & \colhead{Filter} & \colhead{Exp.Time} & \colhead{FWHM} & \colhead{Number of} \\
& & \colhead{(s)} & \colhead{($''$)} & \colhead{fields}
}
\startdata
M15 & $g^{'}$ & 135 & 0.8 & 9 \\
 & $r^{'}$ & 170 & 0.8 & 9 \\
 & $i^{'}$ & 260 & 0.8 & 9 \\
M30 & $g^{'}$ & 110 & 0.7 & 8 \\
 & $r^{'}$ & 125 & 0.7 & 8 \\
 & $i^{'}$ & 160 & 0.6 & 8 \\
M53 / NGC 5053 & $g^{'}$ & 90 & 1.0 & 9 \\
 & $r^{'}$ & 180 & 1.0 & 9 \\
 & $i^{'}$ & 480 & 0.9 & 9 \\
NGC 5466 & $g^{'}$ & 90 & 0.9 & 9 \\
 & $r^{'}$ & 180 & 0.8 & 9 \\
 & $i^{'}$ & 480 & 0.7 & 9 \\
\enddata
\end{deluxetable}
%
\begin{deluxetable}{ccc}
\tabletypesize{\scriptsize}
\tablewidth{0pt}
\tablecaption{The coefficients $a$ and $b$ for a rotational transformation of the 
$(g^{'}-r^{'},r^{'}-i^{'})$  plane to $(c_1, c_2)$ plane, for each cluster.\label{parameter}}
\tablehead{
\colhead{Target} & \colhead{a} & \colhead{b}
}
\startdata
M15 & 0.905 & 0.425 \\
M30 & 0.909 & 0.417 \\
M53 / NGC 5053 & 0.909 & 0.416 \\
NGC 5466 & 0.912 & 0.411\\
\enddata
\end{deluxetable}

\end{document}